\documentclass[12pt]{article}
\usepackage{graphicx}
\usepackage{amssymb}
\usepackage{amsmath}
\usepackage{bm}
\usepackage{url}

\newcommand{\bear}{\begin{array}}
\newcommand {\eear}{\end{array}}
\newcommand{\bea}{\begin{eqnarray}}   
\newcommand{\eea}{\end{eqnarray}}
\newcommand{\beq}{\begin{eqnarray}}   
\newcommand{\eeq}{\end{eqnarray}}
\newcommand{\bef}{\begin{figure}}  \newcommand 
{\eef}{\end{figure}}
\newcommand{\bec}{\begin{center}}  \newcommand 
{\eec}{\end{center}}

\newcommand{\Slash}[1]{{\ooalign{\hfil/\hfil\crcr$#1$}}}
\newcommand{\1}{\mbox{1}\hspace{-0.25em}\mbox{l}}

\begin{document}

\begin{titlepage}

\begin{flushright}
IPMU 12-0065 \\
TU-903
\end{flushright}

\vskip 1.35cm
\begin{center}

{\large 
{\bf 
Reevaluation of Neutron Electric
Dipole Moment \\with QCD Sum Rules
}
}

\vskip 1.2cm

Junji Hisano$^{a,b}$,
Jeong Yong Lee$^a$,
Natsumi Nagata$^{a,c}$,
Yasuhiro Shimizu$^{d,e}$

\vskip 0.4cm
{\it $^a$Department of Physics,
Nagoya University, Nagoya 464-8602, Japan}\\
{\it $^b$IPMU, TODIAS,
University of Tokyo, Kashiwa 277-8568, Japan}\\
{\it $^c$Department of Physics, 
University of Tokyo, Tokyo 113-0033, Japan}\\
{\it $^d$Department of Physics, Tohoku University, Sendai, 980-8578
 Japan}\\
{\it $^e$IIAIR, Tohoku University, Sendai, 980-8578 Japan}

\date{\today}

\begin{abstract} 

  We study the neutron electric dipole moment in the presence of the
  CP-violating operators up to the dimension five in terms of the QCD
  sum rules. It is found that the OPE calculation is robust when
  exploiting a particular interpolating field for neutron, while there
  exist some uncertainties on the phenomenological side. By using input
  parameters obtained from the lattice calculation, we derive a
  conservative limit for the contributions of the CP violating
  operators. We also show the detail of the derivation of the sum
  rules.

\end{abstract}

\end{center}
\end{titlepage}

\section{Introduction}

A variety of experimental efforts \cite{Nakamura:2010zzi} has
precisely determined the elements of the Cabibbo-Kobayashi-Maskawa
(CKM) matrix \cite{Cabibbo:1963yz,Kobayashi:1973fv}, which is a source
of CP violation in the Standard Model (SM). All of CP-violating
processes observed ever are well-explained in terms of the single
physical phase in the CKM matrix. The SM, which is based on the
$SU(3)_C\times SU(2)_L\times U(1)_Y$ gauge symmetry, allows another
CP-violating interaction: the $\theta$ term in the QCD sector. The
CP-violating phenomena caused by the interaction are, however, quite
different from those induced by the CKM phase; the QCD $\theta$ term gives
rise to the CP violation in the flavor-conserving processes, while the CKM
phase induces the CP violation in the flavor-changing ones.
Furthermore, TeV-scale physics beyond the SM, such as the Minimal
Supersymmetric Standard Model (MSSM), often provides other
CP-violating sources. In fact, additional CP-violating interactions
are necessary from the cosmological point of view, since the observed
baryon asymmetry in the Universe is not explained within the SM
interactions. 

The electric dipole moment (EDM) of neutron is one of the physical
quantities that are quite sensitive to the CP violation in the
flavor-conserving interaction. Since there has been no experimental
evidence for its existence so far, a severe constraint is imposed on the
CP-violating interactions. The currently most stringent limit for the
neutron EDM is given by the Institut Laue-Langevin (ILL) experiment
\cite{Baker:2006ts}: 
\begin{equation}
 |d_n|<2.9\times 10^{-26}~e~{\rm cm}~~~~~(90\%~{\rm C.L.})~.
\end{equation}
Moreover, several experimental projects which use ultra cold neutrons
are now under development and expected to have much improved
sensitivities. For example, the nEDM collaboration at the Paul Scherrer
Institute (PSI) \cite{Bodek:2008gr} plans to deliver a sensitivity of
$\sim 5\times 10^{-27}~e~{\rm cm}$, and eventually to reach into the
regime of $10^{-28}~e~{\rm cm}$. Similar sensitivities are expected to
be achieved by the nEDM Collaboration at the Spallation Neutron Source
(SNS) at the U.S. Oak Ridge National Laboratory \cite{Beck:2011gw}, the
CryoEDM experiment \cite{Balashov:2007zj}, the NOP Collaboration at
J-PARC \cite{Arimoto}, and the experiment at KEK-RCNP-TRIUMF \cite{Masuda}. 
Such high sensitivities provide an opportunity to probe the
flavor-conserving CP-violating interactions in the TeV-scale physics
beyond the SM. Furthermore, we may probe the flavor violation in the new
physic indirectly. Even if the new flavor-violating interactions
are introduced in the new physics, the relative CP phase between them
and the CKM matrix may contribute to the EDM \cite{Hisano:2006mj}. 

In order to translate the experimental limits for the neutron EDM into
constraints on the CP violation on the Lagrangian at parton level, one
needs to obtain a relation between these two quantities. There are
some attempts to derive the relation based on the naive dimensional
analysis, the chiral perturbation theory, and the QCD sum rules, though
they are considered to have large uncertainties. It is ultimately
desired that the lattice QCD simulation would evaluate it in future.
There has been discussion of evaluation of neutron EDM with lattice
simulation \cite{hep-lat/0505022}.

In this work, we evaluate the neutron EDM with the QCD sum rules
\cite{ITEP-73-1978}, including the CP-violating
operators up to the dimension five.  It is considered that the QCD sum
rules allow us to derive the relation more systematically than the
naive dimensional analysis and the chiral perturbation
theory \cite{Hisano:2004tf}. 
Similar attempts have been already made in the previous works, {\it
  e.g.}, in a series of papers by M. Pospelov and A. Ritz
\cite{hep-ph/9904483, hep-ph/0010037} and references
therein. We also derive the sum rules for the neutron EDM, while we
use the lattice QCD simulation result for the low-energy constant in
the numerical evaluation of the neutron EDM.  It is found that this
gives more conservative estimate than carrying out all of the
evaluation within the framework of the QCD sum rules. This approach
provides a way of eliminating theoretical errors from the calculation,
while there still remains uncertainty resulting from the QCD sum rule
technique itself.

This paper is organized as follows. In
Sec.~\ref{sec:effective_lagrangian}, we review the CP-violating
interactions at parton level up to the dimension five.  From
Sec.~\ref{phenomenology}, the analysis of the neutron EDM with the QCD
sum rules starts. In Sec.~\ref{phenomenology} we discuss
phenomenological aspects of the correlator of the interpolating field
to neutron, and, in Sec.~\ref{Neutron interpolating field}, show the
properties of the neutron-interpolating field. In
Sec.~\ref{sec:Quark_propagators}, the quark propagators are
derived on the CP-violating and electromagnetic background. They are
used to evaluate the operator product expansion (OPE) for the
correlator in Sec.~\ref{sec:OPEanalysis}. The sum rules for the
neutron EDM are derived in Sec.~\ref{sec:QCD_Sum_Rules}. We found that
there is a difference between results in Refs.~\cite{hep-ph/9904483,hep-ph/0010037} and ours.  In Sec.~\ref{sec:lambda} 
we extract the low-energy constant from the lattice QCD simulation
result. In Sec.~\ref{sec:result} our numerical results for the neutron
EDM are derived. In Sec.~\ref{sec:pq} the neutron EDM is discussed
assuming the Peccei-Quinn symmetry solves the strong CP problem
\cite{Peccei:1977hh}.  Section~\ref{sec:coclusion} is devoted to
conclusion and discussion. 

In Appendix, we show some useful formulae to derive the quark
condensates in the CP-violating background.  In
Appendix~\ref{sec:quark_condensates}, we estimate the effect of the
CP-violating interactions on the generic quark bi-linear condensate
$\langle 0| \bar{q}\Gamma q |0\rangle$, with $\Gamma$ a $4\times 4$
constant matrix, as well as on the quark and gluon background fields. 
In Appendix~\ref{eom} validity of usage of the
classical equations of motion of quarks in evaluation of the quark condensates
is discussed.  In Appendix~\ref{wilson_line} the Wilson-line operators
for the quark fields are discussed in the Fock-Schwinger gauge.

\section{Effective Lagrangian}
\label{sec:effective_lagrangian}

Let us first express the flavor-conserving CP-violating terms in the
low-energy effective Lagrangian for the system consisting of light quarks
and gluon. We include all of the CP-violating operators up to the dimension
five:
\begin{eqnarray}
 {\cal L}\Slash{_{CP}}=&-&\sum_{q=u,d,s}m_q \bar{q}i\theta_q\gamma_5q +
  \theta_G \frac{\alpha_s}{8\pi}G^A_{\mu\nu}\tilde{G}^{A\mu\nu}\nonumber
  \\
&-&\frac{i}{2}\sum_{q=u,d,s}d_q\bar{q}(F\cdot\sigma)\gamma_5q
-\frac{i}{2}\sum_{q=u,d,s}\tilde{d}_q\bar{q}g_s(G\cdot\sigma)\gamma_5q
~.
\label{Lagrangian}
\end{eqnarray}
Here, $m_q$ represents the quark masses, $F_{\mu\nu}$ and
$G^A_{\mu\nu}$ are the electromagnetic and gluon field strength
tensors, $g_s$ is the strong coupling constant
($\alpha_s=g_s^2/4\pi)$, $F\cdot\sigma\equiv F_{\mu\nu}\sigma^{\mu\nu}$,
$G\cdot\sigma\equiv G^A_{\mu\nu}\sigma^{\mu\nu}T^A$, and
$\tilde{G}^A_{\mu\nu}\equiv
\frac{1}{2}\epsilon_{\mu\nu\rho\sigma}G^{A\rho\sigma}$ with
$\epsilon^{0123}=+1$. $T^A$ denotes the generators in the $SU(3)_C$
algebra.  The second, third and forth terms in Eq.~(\ref{Lagrangian})
are called the effective QCD $\theta$ term, the electric and
chromoelectric dipole moments (CEDMs) for quarks, respectively. The
EDMs and CEDMs for quarks are dimension-five operators, and they
are sensitive to the TeV-scale physics beyond the SM.  The
coefficients of the CP-violating operators, $\theta_q$, $\theta_G$,
$d_q$, and $\tilde{d}_q$, are all assumed to be quite small, and we
keep only the terms up to the first order of these parameters.  

The first two terms in Eq.~(\ref{Lagrangian}) are mutually related by
the chiral rotation. 
Consider the following infinitesimal chiral rotation:
\begin{equation}
 q\to q^{\prime}=\left(
1-i\epsilon\rho_q\gamma_5
\right)q~,
\label{chiral_transform}
\end{equation}
where $\epsilon$ is an infinitesimal real constant and $\rho_q$ are
certain parameters for each quark. The Noether current associated with
the transformation is given as
\begin{equation}
 J_{5\mu}=\sum_{q=u,d,s}\rho_q\bar{q}\gamma_{\mu}\gamma_5 q~.
\end{equation}
The divergence of this current does not vanish. Instead,
\begin{eqnarray}
 \partial^\mu J_{5\mu}&=&
\frac{\alpha_s}{4\pi}(\sum_{q}\rho_q)G^A_{\mu\nu}\tilde{G}^{A\mu\nu}
+\sum_{q}2im_q\rho_q\bar{q}\gamma_5 (
1+i\theta_q \gamma_5 )q \nonumber \\
&&-\sum_{q}\rho_q [d_q\bar{q}(F\cdot\sigma)q+
\tilde{d}_q\bar{q}g_s(G\cdot\sigma)q]
~.
\label{axial_anomaly}
\end{eqnarray}
Hereafter we choose $\rho_q$ as
\begin{equation}
 \rho_q=\theta_q/\theta_Q,~~~~~~\theta_Q\equiv\sum_{q=u,d,s}\theta_q~.
\end{equation}
Then, if we take the infinitesimal parameter in
Eq.~(\ref{chiral_transform}) as $\epsilon=\theta_Q/2$, the Lagrangian in
Eq.~(\ref{Lagrangian}) varies by  
\begin{eqnarray}
 \delta{\cal L} &=& \partial^\mu J_{5\mu}\cdot \frac{\theta_Q}{2}
  \nonumber \\
&=&  \sum_{q}m_q\bar{q}i\theta_q\gamma_5 q
+\theta_Q \frac{\alpha_s}{8\pi}G^a_{\mu\nu}\tilde{G}^{a\mu\nu}~,
\label{delta_L}
\end{eqnarray}
which implies that
\begin{equation}
 {\cal L}\Slash{_{CP}}\to {\cal L}^{\prime}\Slash{_{CP}}
=\bar{\theta}\frac{\alpha_s}{8\pi}G^A_{\mu\nu}\tilde{G}^{A\mu\nu}
-\frac{i}{2}\sum_{q=u,d,s}d_q\bar{q}(F\cdot\sigma)\gamma_5q
-\frac{i}{2}\sum_{q=u,d,s}\tilde{d}_q\bar{q}g_s(G\cdot\sigma)\gamma_5q
~,
\end{equation}
where $\bar{\theta}=\theta_G+\theta_Q$.

Therefore, it is found that the $\gamma_5$-mass terms are always reduced
to the ordinary ones, and it is $\bar{\theta}$ that is regarded as a
physical parameter. Of course, one may in turn rotate out the $\theta$ term
into the imaginary mass term through an appropriate chiral rotation. 

In addition, there remains still some arbitrariness in the quark mass
phases $\theta_q$, since they are redefined into another through an
$SU(3)$ chiral rotation. In this article, we choose an appropriate set
of $\theta_q$ so that the choice significantly reduces the
CP-violating contribution to the vacuum expectation values (VEVs) of
quark bi-linear.  We take the condition in Ref.~\cite{Crewther:1979pi}
to determine $\theta_q$, that is, after the $\theta$ term rotated into
the $\gamma_5$-mass term, the following relation should be satisfied:
\begin{equation}
 \langle \Omega_{\tiny \Slash{\rm CP}} |{\cal L}_{\tiny \Slash{\rm CP}}
  |M^A\rangle=0~,~~~~(M^A=\pi,~K,~\eta)~. 
  \label{tadpole_conditions}
\end{equation}
The above condition is evaluated by using the partially conserved
axial-vector current (PCAC) relations. In the current case, it is
sufficient to examine the conditions for $\pi^0$ and $\eta^0$. By using
the PCAC relations, one may readily deduce the conditions for 
the CP-violating parameters from Eq.~(\ref{tadpole_conditions}):
\begin{eqnarray}
 \bar{\theta}(m_u\rho_u-m_d\rho_d)&=&\frac{1}{2}m_0^2
  (\tilde{d}_u-\tilde{d}_d) ~, \nonumber \\
 \bar{\theta}(m_u\rho_u+m_d\rho_d-2m_s\rho_s)&=&\frac{1}{2}m_0^2
  (\tilde{d}_u+\tilde{d}_d-2\tilde{d}_s) ~.
\label{tadpole_conditions2}
\end{eqnarray}
In the calculation we parametrize the
condensate $\langle \bar{q}g_s(G\cdot\sigma)q\rangle$ as \cite{Belyaev:1982sa}
\begin{equation}
 \langle \bar{q}g_s(G\cdot\sigma)q\rangle =-m_0^2 \langle \bar{q}q\rangle~.
\end{equation}
With the relation $\sum_{_q}\rho_q=1$, we then determine the quark mass
phases as follows:
\begin{align}
 \rho_u =&\frac{m_*}{m_u}\biggl[
1+\frac{m_0^2}{2\bar{\theta}}\biggl\{
\frac{\tilde{d}_u-\tilde{d}_d}{m_d}+
\frac{\tilde{d}_u-\tilde{d}_s}{m_s}
\biggr\}
\biggr]~, \nonumber \\
 \rho_d =&\frac{m_*}{m_d}\biggl[
1+\frac{m_0^2}{2\bar{\theta}}\biggl\{
\frac{\tilde{d}_d-\tilde{d}_u}{m_u}+
\frac{\tilde{d}_d-\tilde{d}_s}{m_s}
\biggr\}
\biggr]~, \nonumber \\
 \rho_s =&\frac{m_*}{m_s}\biggl[
1+\frac{m_0^2}{2\bar{\theta}}\biggl\{
\frac{\tilde{d}_s-\tilde{d}_u}{m_u}+
\frac{\tilde{d}_s-\tilde{d}_d}{m_d}
\biggr\}
\biggr]~,
\label{quark_mass_phases}
\end{align}
where
\begin{equation}
 m_*\equiv\frac{m_um_dm_s}{m_um_d+m_dm_s+m_um_s}~.
\end{equation}

\section{Phenomenological behavior of  correlator}
\label{phenomenology}

The QCD sum rules are based on an analysis of the correlator of
interpolating fields\footnote{
There are many review articles about the QCD sum
rules. For example, see Refs.~\cite{review}. 
}.
In the method, OPE allows one to consistently separate the long- and
short-distance contributions to the correlator, and the long-distance
contributions are evaluated by condensations of quarks and gluon. By
comparing the evaluated correlator with the phenomenological model,
the properties for the low-lying parts of the hadronic spectrum are
derived. The Borel transformation is applied to the correlator
there. In this section, we first discuss the phenomenological model
for the correlator.

In the present case, the interpolating field must have the same
quantum numbers as those of neutron, and it is denoted by $\eta_n(x)$
hereafter. On a background with CP-violating sources, the matrix
element of the interpolating field between the vacuum and the
one-particle neutron state is given as
\begin{equation}
 \langle \Omega{\Slash{_{CP}}} |
  \eta_n(x)|N{\Slash{_{CP}}}(\bm{p},s)\rangle =
  {Z^{\frac{1}{2}}_{n,}}{\Slash{_{CP}}}\cdot
  u_{n,}{\Slash{_{CP}}}(\bm{p},s) ~e^{-ip\cdot x}~,
\label{matrix_int}
\end{equation}
where $| \Omega{\Slash{_{CP}}}\rangle$ and
$|N{\Slash{_{CP}}}(\bm{p},s)\rangle$ indicate the vacuum and the one-particle
neutron state on the CP-violating background, respectively. The spinor
$u_{n,}{\Slash{_{CP}}}(\bm{p},s)$ is on-shell neutron wave function which
satisfies the Dirac equation:
\begin{equation}
(\Slash{p}-{m}_{n,}\Slash{_{CP}}\cdot e^{-i\alpha_n
 \gamma_5})u_{n,}{\Slash{_{CP}}}(\bm{p},s)=0 ~.
\label{Dirac_eq}
\end{equation}
Here we include a phase factor $e^{-i\alpha_n \gamma_5}$ into the mass
term, which in general might appear as CP is broken in the vacuum.
Since ${Z^{\frac{1}{2}}_{n,}}{\Slash{_{CP}}}$ and
${m}_{n,}\Slash{_{CP}}$ are both even in terms of the CP-violating
parameters \cite{hep-lat/0505022}, up to the first order of them, 
\begin{equation}
 {Z^{\frac{1}{2}}_{n,}}{\Slash{_{CP}}}=Z^{\frac{1}{2}}_n,~~~~~~
{m}_{n,}\Slash{_{CP}}=m_n~,
\end{equation}
where $m_n$ is the mass of neutron and $\lambda_n\equiv
Z^{\frac{1}{2}}_n$ is the coupling between the physical neutron state
and the interpolating field without CP-violating sources.
Then the solution of Eq.~(\ref{Dirac_eq}) turns out to be
\begin{equation}
 u_{n,}{\Slash{_{CP}}}(\bm{p},s)=e^{\frac{i}{2}\alpha_n
  \gamma_5}u_n(\bm{p}, s)~,
\end{equation}
with $u_n(\bm{p}, s)$ an ordinary spinor wave function which satisfies
$(\Slash{p}-m_n)u_n(\bm{p}, s)=0$. As a result Eq.~(\ref{matrix_int})
leads to
\begin{equation}
 \langle \Omega{\Slash{_{CP}}} |
  \eta_n(x)|N{\Slash{_{CP}}}(\bm{p},s)\rangle =
\lambda_n
e^{\frac{i}{2}\alpha_n
  \gamma_5}~u_n(\bm{p}, s)~e^{-ip\cdot x}~.
\label{matrix_interpolating}
\end{equation}
The low-energy constant $\lambda_n$ is to be determined later.

Now we analyze the correlator of the interpolating fields from the
phenomenological viewpoint. It is defined as
\begin{equation}
 \Pi(q)\equiv i\int d^4 x ~e^{iq\cdot x} ~\langle \Omega{\Slash{_{CP}}}
  |T\{\eta_n(x)\bar{\eta}_n(0)\} |\Omega{\Slash{_{CP}}}\rangle_{F}~,
\label{correlator}
\end{equation}
where the subscript $F$ implies that the correlator
is evaluated on an electromagnetic field background. Our goal is to
extract the EDM of neutron from the correlator. 
The phase factor in Eq.~(\ref{matrix_interpolating}), however, causes mixture
between electric and magnetic dipole moment structures and makes it difficult
to pick out only the EDM from the QCD sum rules. So we first examine the
Lorentz structures of the correlator and select a term independent of
the phase $\alpha_n$, {\it i.e.}, chiral invariant. As discussed in
Ref.~\cite{hep-ph/9904483}, up to the leading order on
the background electromagnetic field, the correlator $\Pi(q)$ is
estimated by inserting an effective vertex such as
\begin{equation}
 {\cal L}_n=-\frac{i}{2} d_n \bar{N}(F\cdot\sigma)\gamma_5 N
=\frac{d_n}{2}\bar{N}\tilde{F}\cdot\sigma N~.
\end{equation}
Here, $N(\equiv N(x))$ denotes the renormalized neutron field which is
approximately equivalent to $\lambda^{-1}_ne^{-i\alpha_n \gamma_5/2}
\eta_n(x)$, and $d_n$ is the EDM of neutron. A similar procedure to
those in Ref.~\cite{hep-ph/9904483} shows that terms with an
odd number of Dirac matrices are independent of the phase factor
$\alpha_n$, and furthermore, those proportional to $\{\tilde{F}\cdot\sigma ,
\Slash{q}\}$ are the unique choice in this case. Therefore we only focus
on such terms in the following calculation. Then, the phenomenological
expression of the correlator is found to be\footnote{
In the published versions of Refs.~\cite{hep-ph/9904483, hep-ph/0010037}
the coefficient of the double pole in Eq.~(\ref{phenomenology_f}) is
different from ours by a factor of two, while that in the revised arXiv
versions are consistent with ours. 
}.
\begin{equation}
 \Pi^{(\rm phen)}(q)=\frac{1}{2}f(q^2)\{\tilde{F}\cdot\sigma,
  \Slash{q}\}+\dots~,
\label{corr_phen}
\end{equation}
where dots indicate terms with other Lorentz structures and 
\begin{equation}
 f(q^2)=\left(
\frac{\lambda^2_n d_n m_n}{(q^2-m_n^2)^2}+\frac{A(q^2)}{q^2-m_n^2}+B(q^2)
\right)
\label{phenomenology_f}
\end{equation}
with $A(q^2)$ and $B(q^2)$ functions which have no pole at $q^2=m_n^2$.
As noted in Ref.~\cite{hep-ph/9904483}, since we are 
effectively dealing with a three-point function, it might be
inconsistent to parametrize the continuum contribution in terms of a
usual ansatz for the spectral function with a certain threshold in the
QCD sum rules.  We just neglect the contribution with expecting its
significance to be small enough. Furthermore, we assume that the
function $A(q^2)$ has little dependence on $q^2$, and regard it as a
constant when we conduct the Borel transformation.

\section{Neutron-interpolating field}
\label{Neutron interpolating field}

In this section we give discussion on choice of the
neutron-interpolating field which we use for the QCD sum rule
calculation. The field must have the same quantum numbers as
neutron. The most general interpolator for neutron on the ordinary
CP-even background is parametrized as
\begin{equation}
 \eta_n(x)=j_1(x)+\beta j_2(x)~,
\end{equation}
where
\begin{equation}
 j_1(x)=2\epsilon_{abc}\left(
d^{T}_a(x)C\gamma_5 u_b(x)
\right)d_c(x)~,
\label{j1}
\end{equation}
and
\begin{equation}
  j_2(x)=2\epsilon_{abc}\left(
d^{T}_a(x)C u_b(x)
\right)\gamma_5 d_c(x)~.
\label{j2}
\end{equation}
Here the subscripts, $a,b,c$, denote the color indices and $C$ is the charge
conjugation matrix. The interpolator $j_1(x)$ is often used in lattice
simulations. While $j_2(x)$ vanishes in the non-relativistic limit, it
should be included to the whole interpolating field since we deal with
light quarks. The unphysical parameter $\beta$ is to be fixed later so that 
the calculation is transparent.

When the calculation is carried out on the CP-violating background,
however, the interpolating fields include additional components. This
point is easily understood when one considers the chiral rotation
discussed in Sec.~\ref{sec:effective_lagrangian}. As we have seen in
Sec.~\ref{sec:effective_lagrangian}, the chiral rotation
(\ref{chiral_transform}) transforms the Lagrangian ${\cal L}$ into
another. The same transformation, in turn, changes the interpolators
$j_1(x)$ and $j_2(x)$ into other ones as
\begin{eqnarray}
 j_1(x)&\to& j_1(x)-i\epsilon [(\rho_u+\rho_d)i_1(x)+\rho_di_2(x)]~,
\nonumber \\
 j_2(x)&\to& j_2(x)-i\epsilon [(\rho_u+\rho_d)i_2(x)+\rho_di_1(x)]~,
\end{eqnarray}
where
\begin{eqnarray}
 i_1(x)&=&2\epsilon_{abc}(d^T_a(x)Cu_b(x))d_c(x)~, \nonumber \\
 i_2(x)&=&2\epsilon_{abc}(d^T_a(x)C\gamma_5u_b(x))\gamma_5d_c(x)~.
\end{eqnarray}
Therefore, with generic CP-violating terms as in Eq.~(\ref{Lagrangian}),
the interpolators acquire mixing terms as
\begin{eqnarray}
 \tilde{j}_1(x)&=&j_1(x)+i\epsilon i_1(x)+i \delta i_2(x)~,\nonumber \\
 \tilde{j}_2(x)&=&j_2(x)+i\epsilon i_2(x)+i \delta i_1(x)~,\nonumber \\
 \tilde{i}_1(x)&=&i_1(x)+i\epsilon j_1(x)+i \delta j_2(x)~,\nonumber \\
 \tilde{i}_2(x)&=&i_2(x)+i\epsilon j_2(x)+i \delta j_1(x)~,
\end{eqnarray}
with $\epsilon$ and $\delta$ the small constants which are
suppressed by the CP-violating parameters. Furthermore, the above
expressions are rewritten as
\begin{eqnarray}
 \tilde{j}_1(x)&=&(1+i\delta\gamma_5)[j_1(x)+i\epsilon i_1(x)]~,\nonumber \\
 \tilde{j}_2(x)&=&(1+i\delta\gamma_5)[j_2(x)+i\epsilon i_2(x)]~,\nonumber \\
 \tilde{i}_1(x)&=&(1+i\delta\gamma_5)[i_1(x)+i\epsilon j_1(x)]~,\nonumber \\
 \tilde{i}_2(x)&=&(1+i\delta\gamma_5)[i_2(x)+i\epsilon j_2(x)]~,
\end{eqnarray}
because
\begin{eqnarray}
 i_1(x)&=&\gamma_5 j_2(x)~, \nonumber \\
 i_2(x)&=&\gamma_5 j_1(x)~.
\end{eqnarray}
Now that we concentrate on the chiral-invariant structure in the
correlator of the  neutron-interpolating field as discussed in
Sec.~\ref{phenomenology}, the infinitesimal chiral rotation factor
$(1+i\delta\gamma_5)$ is ignorable. After all, the neutron-interpolating
field which we deal with has a following structure:
\begin{equation}
 \eta_n(x)=j_1(x)+\beta j_2(x)+i\epsilon[i_1(x)+\beta i_2(x)]~. 
\end{equation}

The small constant $\epsilon$ is determined by the condition that the
interpolating field $\eta_n(x)$ has a vanishing correlator with the
current $\xi_n(x)$ defined as follows:
\begin{equation}
 \xi_n(x)=i_1(x)+\beta i_2(x)+i\epsilon[j_1(x)+\beta j_2(x)]~. 
\end{equation}
In what follows, however, we sweep away the contribution of the mixture
terms in the interpolating field by choosing an appropriate value for
the parameter $\beta$. In the subsequent sections, we calculate the
correlator of $\eta_n(x)$ by using the OPE method. The correlator is
expressed by the sum of the correlators for each component interpolator
as
\begin{eqnarray}
 \langle \Omega{\Slash{_\mathrm{CP}}}|T\{\eta_n(x)\bar{\eta}_n(0)\}|
  \Omega{\Slash{_\mathrm{CP}}}\rangle_F 
&=&
  \langle j_1,\bar{j}_1\rangle +\beta[\langle j_1,\bar{j}_2\rangle +
\langle j_2,\bar{j}_1\rangle]
+ \beta^2 \langle j_2,\bar{j}_2\rangle \nonumber \\
&& + i\epsilon [\gamma_5\langle j_2,\bar{j}_1\rangle +\langle
 j_1,\bar{j}_2\rangle\gamma_5] \nonumber \\
&& + i\epsilon\beta [\{\langle j_1, \bar{j}_1\rangle, \gamma_5\}+ 
\{\langle j_2, \bar{j}_2 \rangle, \gamma_5 \} ] \nonumber \\
&& + i\epsilon\beta^2 [\gamma_5\langle j_1, \bar{j}_2\rangle +\langle j_2,
 \bar{j}_1 \rangle\gamma_5]~,
\end{eqnarray}
where
\begin{equation}
 \langle a, \bar{b}\rangle \equiv 
\langle \Omega{\Slash{_\mathrm{CP}}}|T\{a(x)\bar{b}(0)\}|
  \Omega{\Slash{_\mathrm{CP}}}\rangle_F~.
\end{equation}
As discussed in Sec.~\ref{phenomenology}, we focus on parts of the
correlators which have the Lorentz structures with an odd number of gamma
matrices. Such terms anti-commute with $\gamma_5$. Thus, in this case,
the above expression leads to
\begin{eqnarray}
  \langle \Omega{\Slash{_\mathrm{CP}}}|T\{\eta_n(x)\bar{\eta}_n(0)\}|
  \Omega{\Slash{_\mathrm{CP}}}\rangle_F |_{\gamma~\rm odd}
&=&
  \langle j_1,\bar{j}_1\rangle +\beta[\langle j_1,\bar{j}_2\rangle +
\langle j_2,\bar{j}_1\rangle]
+ \beta^2 \langle j_2,\bar{j}_2\rangle \nonumber \\
&&+ i\epsilon(1-\beta^2) [\langle j_1,\bar{j}_2\rangle
-\langle j_2,\bar{j}_1\rangle ]\gamma_5 ~.
\end{eqnarray}
This equation shows that the mixing terms in the interpolating field do
not affect the correlator if one sets $\beta$ to be $\pm 1$. 
As we will see later, 
for our calculation, $\beta =+1$ is an appropriate choice since this
choice eliminates the sub-leading terms with infrared logarithm, which
yield ambiguity due to the infrared cutoff\footnote{
The choice of $\beta$ for the sum rules including only the QCD $\theta$
term is discussed in Ref.~\cite{Narison:2008jp}. They argue that
optimal choice of $\beta$ is $-1\leq \beta \leq 0$ rather than 1, which
is consistent with the conventional choices favored from a viewpoint of
evaluation of $\lambda_n$.  
The discussion is, however, not applicable
to the present case since our sum rules contain several unknown
parameters. 
}. With this choice one may simultaneously exclude the contribution
of the mixing terms. 
Thus, we will not calculate such mixing contributions with keeping
in mind that we will finally take $\beta=+1$ when we
derive the QCD sum rules\footnote{
The neutron-interpolating field for $\beta=+1$ is simply expressed as 
\begin{equation}
 \eta_n(x)=\frac{1}{2} \epsilon^{abc}\bigl(
d^T_aC \sigma_{\mu\nu}d_b
\bigr) \sigma^{\mu\nu} \gamma_5  u_c~.
\end{equation}
}. 
That is to say, we deal with the correlator
\begin{eqnarray}
  \langle \Omega{\Slash{_\mathrm{CP}}}|T\{\eta_n(x)\bar{\eta}_n(0)\}|
  \Omega{\Slash{_\mathrm{CP}}}\rangle_F 
=
  \langle j_1,\bar{j}_1\rangle +\beta[\langle j_1,\bar{j}_2\rangle +
\langle j_2,\bar{j}_1\rangle]
+\beta^2\langle j_2,\bar{j}_2\rangle ~,
\label{decomposition}
\end{eqnarray}
and after the computation, we set $\beta=+1$.

%
\section{Quark propagators on the CP-violating background}
\label{sec:Quark_propagators}

When evaluating the correlator (\ref{correlator}) in the OPE, we need to
obtain the quark propagators on the CP-violating background with an
electromagnetic background field $F$.
They are defined as follows:
\begin{equation}
 [S^q_{ab}(x) ]_{\alpha\beta}
\equiv
\langle\Omega{\Slash{_\mathrm{CP}}} |T
\left[q_{a\alpha}(x)\bar{q}_{b\beta}(0) \right]| 
\Omega{\Slash{_\mathrm{CP}}}\rangle_F~, 
\end{equation}
where $\alpha$ and $\beta$ denote spinor indices. 
Expanding the propagators as
\begin{equation}
  \left[S^q_{ab}(x) \right]_{\alpha\beta}
=
 \left[S^{q(0)}_{ab}(x) \right]_{\alpha\beta}+
\chi^q_{a\alpha}(x)\bar{\chi}^q_{b\beta}(0)
+
\left[S^q_{ab}(x) \right]_{\alpha\beta}|_{1~{\rm photon}}+
\left[S^q_{ab}(x) \right]_{\alpha\beta}|_{1~{\rm gluon}}+ \dots~,
\label{propagator}
\end{equation}
we evaluate each term in $x$-space. The first term is the free
propagator, and the second term describes the correlator of the quark
background fields, with $\chi^q_{a\alpha}(x)$ a classical Grassmann field
which indicates the quark background field.
The third and fourth terms represent the
propagators including one photon and gluon, respectively.
In the derivation of 
the quark propagators we use
the classical equations of motion for quark fields given as
\begin{eqnarray}
 i \Slash{D}q &=& m_q(1+i\theta_q  \gamma_5 )q
+\frac{i}{2}\sum_{q=u,d,s}d_q(F\cdot\sigma)\gamma_5q
+\frac{i}{2}\sum_{q=u,d,s}\tilde{d}_qg_s(G\cdot\sigma)\gamma_5q~, \nonumber\\
 -i\bar{q}\overleftarrow{\Slash{D}}&=&m_q\bar{q}(1+i\theta_q\gamma_5)
+\frac{i}{2}\sum_{q=u,d,s}d_q\bar{q}(F\cdot\sigma)\gamma_5
+\frac{i}{2}\sum_{q=u,d,s}\tilde{d}_qg_s\bar{q}(G\cdot\sigma)\gamma_5
~,
\label{equation_of_motion}
\end{eqnarray}
where $D_{\mu}=\partial_{\mu}-ie_qA_\mu -ig_sG^A_\mu T^A$ and
$\bar{q}\overleftarrow{\Slash{D}}=\partial_\mu\bar{q}\gamma^{\mu}
+ie_q\bar{q}\Slash{A}+ig_s\bar{q}\Slash{G}^A T^A$.  The
electromagnetic and gluon fields are denoted as $A_\mu$ and $G^A_\mu$,
respectively, with $e_q$ the quark charge.

The first term in Eq.~(\ref{propagator}), $[S^{q(0)}_{ab}(x)
]_{\alpha\beta}$, is readily evaluated by using the equations of motion
without electromagnetic and gluon background fields. The result is
\begin{equation}
 S^{q(0)}_{ab}(x) =\frac{i\delta_{ab}}{2\pi^2}\frac{\Slash{x}}{(x^2)^2}
-\frac{m_q\delta_{ab}}{4\pi^2x^2}\left(1-i\theta_q\gamma_5
\right)~,
\label{propagator0}
\end{equation}
where we keep the terms up to the first order of the quark mass $m_q$.

Next, we evaluate the third and fourth terms in
Eq.~(\ref{propagator}). These terms again are obtained from the
equations of motion (\ref{equation_of_motion}). In this calculation, it
is convenient to use the Fock-Schwinger gauge \cite{Novikov:1983gd} for both
the electromagnetic and the gluon fields:
\begin{equation}
 x^\mu A_\mu(x)= x^\mu G^A_\mu(x)=0~.
\end{equation}
In this gauge, the fields are expanded by their field strength, such as
\begin{equation}
 A_\mu(x)=
\frac{1}{2\cdot 0 !} x^{\nu} F_{\nu\mu}(0)
+
\frac{1}{3\cdot 1 !} 
x^\alpha x^{\nu} (D_\alpha  F_{\nu\mu}(0))
+
\frac{1}{4\cdot 2 !} 
x^\alpha x^\beta x^{\nu} (D_\alpha D_\beta F_{\nu\mu}(0))
+
\cdots.
\label{Aexpansion}
\end{equation}
By using the expression, the gauge covariant form
of the propagators is obtained as follows:
\begin{eqnarray}
 S^q_{ab}(x)|_{\rm 1~photon}=
&-&\frac{ie_q}{32\pi^2}\delta_{ab}\biggl[
\frac{1}{x^2}\{\Slash{x},F\cdot \sigma\}-im_q(1-i\theta_q\gamma_5)F\cdot
\sigma\log (-\Lambda_{IR}^2 x^2)
\biggr] \nonumber \\
&-&\frac{d_q}{8\pi^2}\delta_{ab}\biggl[
\frac{\Slash{x}\tilde{F}\cdot \sigma\Slash{x}}{(x^2)^2}+
\frac{m_q}{2x^2}\{\Slash{x}, \tilde{F}\cdot \sigma\}
\biggr]~, 
\label{photon_propagator}
\end{eqnarray}
\begin{eqnarray}
 S^q_{ab}(x)|_{\rm 1~gluon}=
&-&\frac{ig_s}{32\pi^2}\biggl[
\frac{1}{x^2}\{\Slash{x},G_{ab}\cdot \sigma\}
-im_q(1-i\theta_q\gamma_5)G_{ab}\cdot
\sigma\log (-\Lambda_{IR}^2 x^2)
\biggr] \nonumber \\
&-&\frac{\tilde{d}_qg_s}{8\pi^2}\biggl[
\frac{\Slash{x}\tilde{G}_{ab}\cdot \sigma\Slash{x}}{(x^2)^2}+
\frac{m_q}{2x^2}\{\Slash{x}, \tilde{G}_{ab}\cdot \sigma\}
\biggr]~,
\label{gluon_propagator}
\end{eqnarray}
with $G_{ab}^{\mu\nu}=G^{A\mu\nu}T^A_{ab}$. Here, a certain infrared
(IR) cutoff $\Lambda_{IR}$ is introduced in logarithmic terms. It was
replaced to the quark masses when deriving the propagator from the
equation motions. However, the contribution to the OPE with small
quark momenta around the quark masses should not be included so that
the IR cutoff is introduced.

Finally, we translate the quark and gluon background
fields into their condensates. Here we just give resultant expressions
for the relations between them. The details of the derivation is
presented in Appendix~\ref{sec:quark_condensates}.

A single quark line, $\chi^q_{a\alpha}(x)\bar{\chi}^q_{b\beta}(0)$, is
related with the quark condensate as
\begin{equation}
 \chi^q_{a\alpha}(x)\bar{\chi}^q_{b\beta}(0)=
\langle \Omega{\Slash{_\mathrm{CP}}}|
q_{a\alpha}(x)\bar{q}_{b\beta}(0) |\Omega{\Slash{_\mathrm{CP}}}
\rangle_F~,
\end{equation}
and it is to be expressed in terms of
$\langle \bar{q}q\rangle$ as follows:
\begin{align}
  \chi^q_{a\alpha}(x)\bar{\chi}^q_{b\beta}(0)
=&-\frac{\delta_{ab}}{12}\left(
1+i\theta_G\rho_q\gamma_5
\right)_{\alpha\beta}
\langle\bar{q}q\rangle
+\frac{i}{48}\delta_{ab}\Slash{x}_{\alpha\beta}m_q
\langle\bar{q}q\rangle \nonumber \\ 
 &-\frac{i}{96}\delta_{ab}\left[\bar{\theta}m_q\rho_q e_q \chi 
+d_q+(\kappa-\frac{1}{2}\xi)e_q\tilde{d}_q
\right]
\{
\tilde{F}\cdot \sigma, \Slash{x}
\}_{\alpha\beta}
\langle\bar{q}q\rangle \nonumber \\
&+\frac{i}{96}m_qe_q \chi \delta_{ab}
\{
F\cdot \sigma, \Slash{x}
\}_{\alpha\beta}
\langle\bar{q}q\rangle \nonumber \\ 
 &-\frac{i}{24}e_q \chi \delta_{ab}
\left(\tilde{F}\cdot\sigma\gamma_5[1+i
\rho_q\theta_G\gamma_5]
\right)_{\alpha\beta}
\langle\bar{q}q\rangle ~.
\end{align}
Here, $\chi,~\kappa$ and $\xi$ are the parameters for the quark
condensates defined as \cite{Ioffe:1983ju}
\begin{equation}
 \langle \bar{q}\sigma_{\mu\nu}q\rangle_F   =e_q \chi
  F_{\mu\nu}\langle \bar{q}q\rangle ,
\end{equation}
\begin{equation}
 g_s \langle \bar{q}G^A_{\mu\nu}T^Aq\rangle_F=e_q \kappa  F_{\mu\nu}
\langle \bar{q}q\rangle~,
\end{equation}
\begin{equation}
 2 g_s \langle \bar{q}\gamma_5 \tilde{G}^A_{\mu\nu}T^Aq\rangle_F
= ie_q\xi  F_{\mu\nu}
\langle \bar{q}q\rangle~.
\end{equation}
Also, in our calculation, we need the interaction part of the quark and
gluon background fields,  
\begin{equation}
 g_s\chi^q_{a\alpha}(x) \bar{\chi}^q_{b\beta}(0)
[G_{\mu\nu}]_{cd}=
\langle g_s q_{a\alpha}(x)[G_{\mu\nu}]_{cd} \bar{q}_{b\beta}(0)
\rangle_{F,{\tiny \Slash{\rm
CP}}}~,
\end{equation}
and it leads to the following equation:
\begin{align}
  g_s\chi^q_{a\alpha}(x) \bar{\chi}^q_{b\beta}(0)
[G_{\mu\nu}]_{cd} =
&-
 \frac{1}{32}(\delta_{ad}\delta_{bc}-\frac{1}{3}\delta_{ab}\delta_{cd})
\langle\bar{q}q\rangle
\nonumber \\
&\times\biggl[
e_q(\kappa F_{\mu\nu}-\frac{i}{2}\xi \tilde{F}_{\mu\nu}\gamma_5)
(1+i\theta_G \rho_q \gamma_5)
 \nonumber \\
&-\frac{i}{4}m_qe_q\Slash{x}(\kappa F_{\mu\nu}+\frac{1}{2}\bar{\theta}
 \rho_q \xi \tilde{F}_{\mu\nu}) \nonumber \\
&-\frac{i}{24}m_q m_0^2 \epsilon_{\mu\nu\rho\sigma} x^\rho \gamma^\sigma
 \gamma_5 -\frac{i}{24}\bar{\theta}m_q\rho_qm_0^2 (x_\mu \gamma_\nu
 \gamma_5 -x_\nu \gamma_\mu \gamma_5) \nonumber \\
&-\frac{1}{12}m_0^2 \sigma_{\mu\nu}
-\frac{i}{12}m_0^2 \theta_G \rho_q \sigma_{\mu\nu} \gamma_5 
\biggr]_{\alpha\beta}~.
\label{chichig}
\end{align}

\section{OPE analysis of the correlator}
\label{sec:OPEanalysis}

\subsection{Leading order}

\begin{figure}[t]
\begin{center}
\includegraphics[height=45mm]{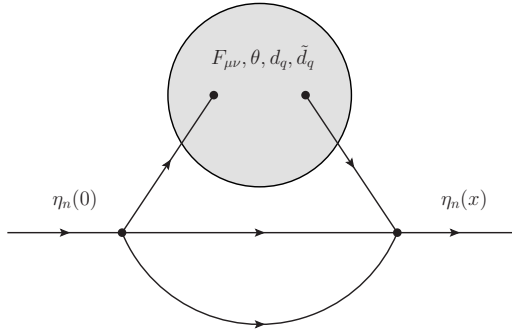}
\caption{Diagram which yields the leading contribution without
 emitting either gluon or photon.}  
\label{leading1}
\end{center}
\end{figure}

\begin{figure}[t]
\begin{center}
\includegraphics[height=45mm]{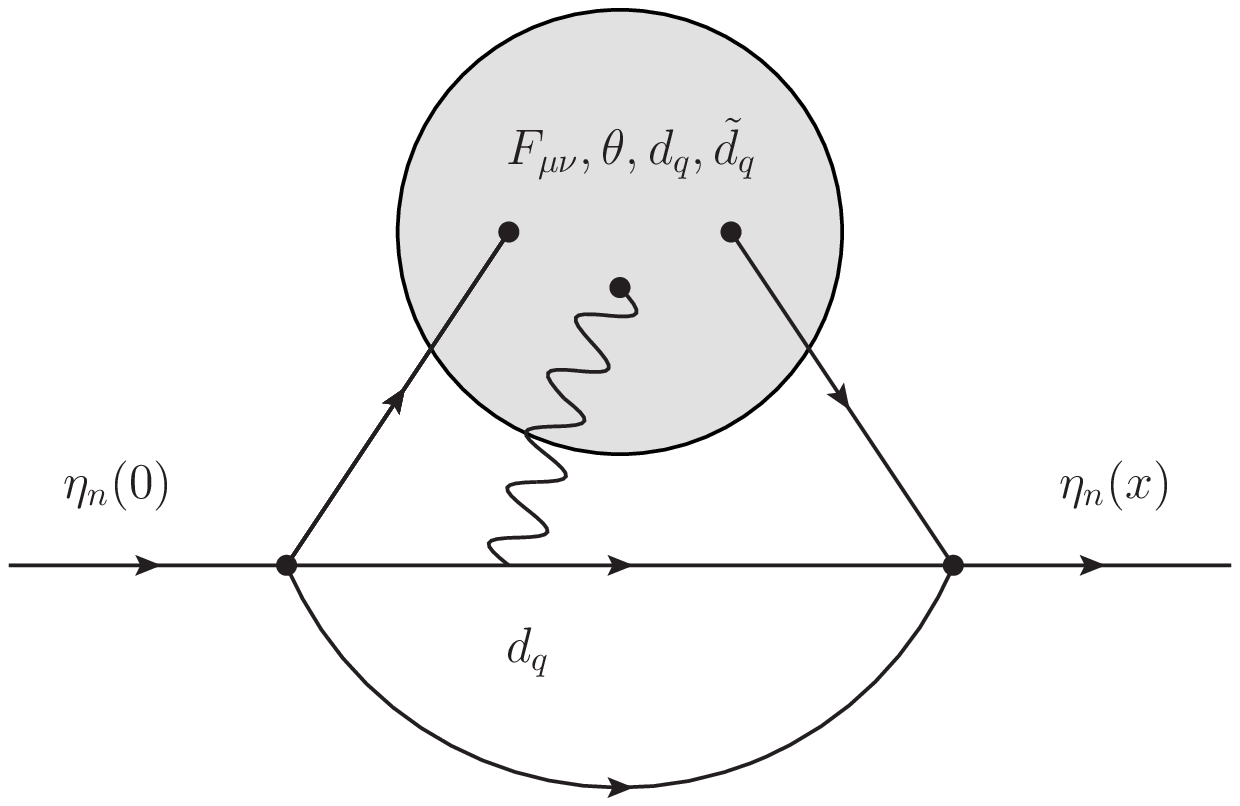}
\caption{Diagram which yields the leading and the next-to-leading order
 contributions with emitting a photon. Those contributions
 vanish when $\beta=+1$.}  
\label{leading2}
\end{center}
\end{figure}

\begin{figure}[t]
\begin{center}
\includegraphics[height=45mm]{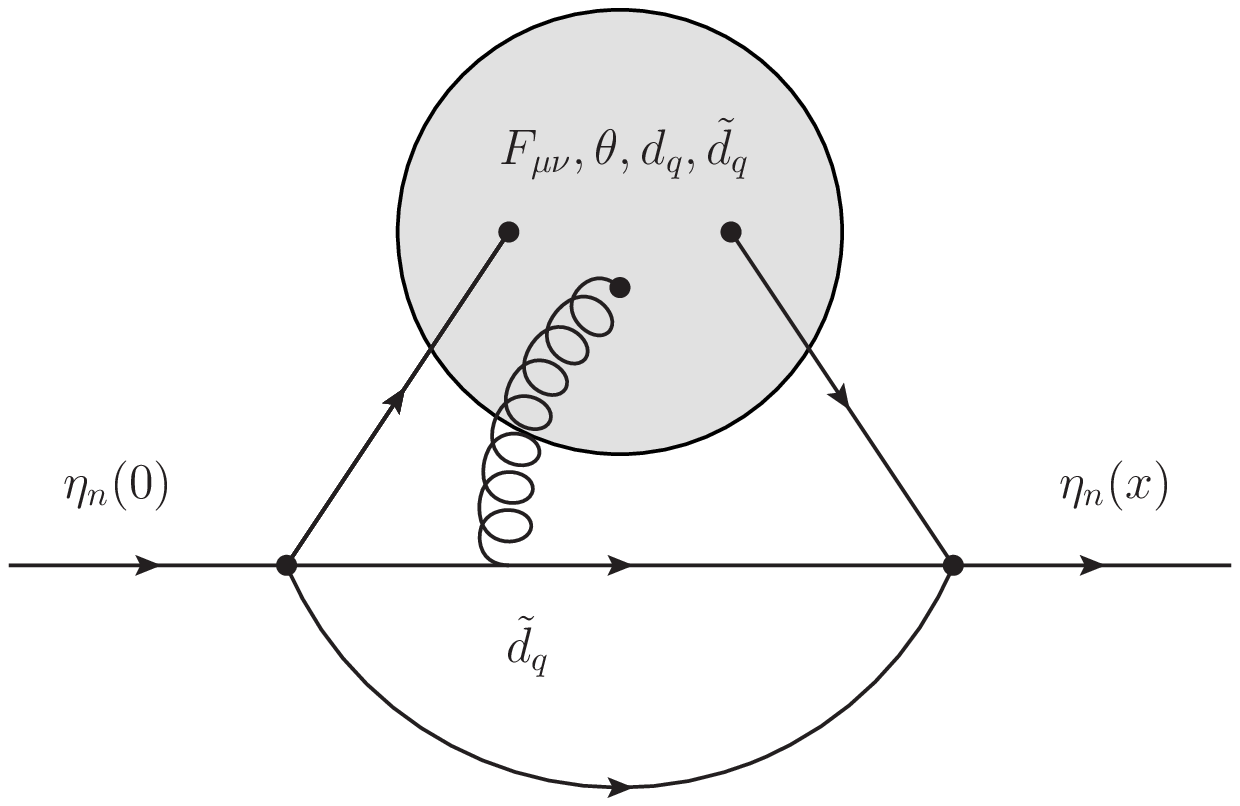}
\caption{Diagram which yields the leading and the next-to-leading order
 contributions with emitting a gluon.  Those contributions
 vanish when $\beta=+1$.}  
\label{leading3}
\end{center}
\end{figure}

Now we calculate the correlation function of the interpolating fields 
$\eta_n(x)$ in terms of the OPE. First, we carry out the
leading-order calculation for the correlator
\begin{equation}
 \Pi(x)=  \langle \Omega{\Slash{_\mathrm{CP}}}|T\{\eta_n(x)\bar{\eta}_n(0)\}|
  \Omega{\Slash{_\mathrm{CP}}}\rangle_F ~.
\end{equation}
As in Eq.~(\ref{decomposition}), this correlator is decomposed into four
correlators. We deal with them inclusively by using the following
notation:
\begin{equation}
  \Pi_{kl}(x)=  
\langle \Omega{\Slash{_\mathrm{CP}}}|T\{j_k(x)\bar{j}_l(0)\}|
  \Omega{\Slash{_\mathrm{CP}}}\rangle_F~,~~~~~~(k,l=1,2) ~.
\end{equation}
In Figs.~\ref{leading1}-\ref{leading3}, the diagrams which contribute to
the correlators are illustrated. We denote each contribution to the
correlators by the upper indices, {\it i.e.}, $\Pi^{(I)}(x)$ or
$\Pi^{(I)}_{kl}(x)$ with $I=1,2,3$. 
From now on, we use the following abbreviation:
\begin{equation}
 \bar{S}^q(x)\equiv CS^{qT}(x)C^{\dagger}~,
\end{equation}
with $C$ the charge conjugation matrix.

Let us begin with evaluating $\Pi^{(1)}_{kl}(x)$. 
Each $\Pi^{(1)}_{kl}(x)$ is expressed in terms of the propagators
$S^q_{ab}(x)$ as
\begin{eqnarray}
 \Pi^{(1)}_{11}(x) &=& 4\epsilon_{abc} \epsilon_{a^\prime b^\prime
  c^\prime} \{-{\rm Tr}[\gamma_5 S^u_{ba^\prime}(x)\gamma_5
  \bar{S}_{ab^\prime }^d(x)
  ]S^d_{cc^\prime}(x) +S^d_{cb^\prime}(x)\gamma_5 \bar{S}_{ba^\prime
  }^u(x) \gamma_5 S_{ac^\prime }^d(x) \} ~,
\nonumber \\
 \Pi^{(1)}_{12}(x) &=& 4\epsilon_{abc} \epsilon_{a^\prime b^\prime
  c^\prime} \{-{\rm Tr}[\gamma_5 S^u_{ba^\prime}(x)
  \bar{S}_{ab^\prime }^d(x)
  ]S^d_{cc^\prime}(x)\gamma_5 +S^d_{cb^\prime}(x)\bar{S}_{ba^\prime
  }^u(x) \gamma_5 S_{ac^\prime }^d(x)\gamma_5 \} ~,
\nonumber \\
 \Pi^{(1)}_{21}(x)&=& 4\epsilon_{abc} \epsilon_{a^\prime b^\prime
  c^\prime} \{-{\rm Tr}[S^u_{ba^\prime}(x)\gamma_5
  \bar{S}_{ab^\prime }^d(x)
  ]\gamma_5S^d_{cc^\prime}(x) +\gamma_5S^d_{cb^\prime}(x)\gamma_5
  \bar{S}_{ba^\prime 
  }^u(x) S_{ac^\prime }^d(x) \} ~,
\nonumber \\
 \Pi^{(1)}_{22}(x) &=& 4\epsilon_{abc} \epsilon_{a^\prime b^\prime
  c^\prime} \{-{\rm Tr}[S^u_{ba^\prime}(x) \bar{S}_{ab^\prime }^d(x)
  ]\gamma_5S^d_{cc^\prime}(x)\gamma_5 +\gamma_5S^d_{cb^\prime}(x)
  \bar{S}_{ba^\prime 
  }^u(x) S_{ac^\prime }^d(x)\gamma_5 \} ~. \nonumber \\
\label{correlator_propagator}
\end{eqnarray}
When the propagators include neither photon nor gluon emitting term, 
the expressions reduce to 
\begin{eqnarray}
 \Pi^{(1)}_{11}(x) &=& 24\{{\rm Tr}[\gamma_5 S^u(x)\gamma_5 \bar{S}^d(x)
  ]S^d(x) +S^d(x)\gamma_5 \bar{S}^u(x) \gamma_5 S^d(x) \} ~,
\nonumber \\
 \Pi^{(1)}_{12}(x) &=& 24\{{\rm Tr}[\gamma_5 S^u(x)\bar{S}^d(x)
  ]S^d(x)\gamma_5 +S^d(x) \bar{S}^u(x) \gamma_5 S^d(x)\gamma_5 \} ~,
\nonumber \\
 \Pi^{(1)}_{21}(x) &=& 24\{{\rm Tr}[ S^u(x)\gamma_5 \bar{S}^d(x)
  ]\gamma_5 S^d(x) +\gamma_5 S^d(x)\gamma_5 \bar{S}^u(x) S^d(x) \} ~,
\nonumber \\
 \Pi^{(1)}_{22}(x) &=& 24\{{\rm Tr}[S^u(x)\bar{S}^d(x)
  ]\gamma_5S^d(x)\gamma_5 +\gamma_5S^d(x)\bar{S}^u(x) S^d(x)\gamma_5 \}~,
\label{corr_prop}
\end{eqnarray}
where
\begin{equation}
 S^q_{ab}(x)=\delta_{ab} S^q (x)~.
\end{equation}

As discussed in Sec.~\ref{phenomenology}, we focus on the terms
proportional to $\{\tilde{F}\cdot\sigma, \Slash{q}\}$. For this reason we
extract only the terms including $\{\tilde{F}\cdot\sigma, \Slash{x}\}$.
They are found to be
\begin{align}
 \Pi^{(1)}_{11}(x) =&
\frac{i}{16\pi^4}\langle \bar{q}q\rangle \frac{1}{(x^2)^3} \{
\tilde{F}\cdot\sigma, \Slash{x}\} \bigl[
(4e_dm_d\rho_d-e_um_u\rho_u)\chi \bar{\theta} \nonumber \\
&+ \chi \{(e_dm_u\rho_u-e_um_d\rho_d)\theta_Q+(e_um_d\rho_u
 -e_dm_u\rho_d )\theta_G\} \nonumber \\
&+(6d_d-d_u) +(\kappa -\frac{1}{2}\xi )
(6e_d\tilde{d}_d -e_u\tilde{d}_u)
\bigr]~, \nonumber \\
 \Pi^{(1)}_{12}(x) =& \Pi^{(1)}_{21}(x) \nonumber \\
=& \frac{i}{16\pi^4}\langle \bar{q}q\rangle \frac{1}{(x^2)^3} \{
\tilde{F}\cdot\sigma, \Slash{x}\} \bigl[
(4e_dm_d\rho_d-e_um_u\rho_u)\chi \bar{\theta} \nonumber \\
&+(2d_d-d_u) +(\kappa -\frac{1}{2}\xi )
(2e_d\tilde{d}_d -e_u\tilde{d}_u)
\bigr]~, \nonumber \\
 \Pi^{(1)}_{22}(x) =&
\frac{i}{16\pi^4}\langle \bar{q}q\rangle \frac{1}{(x^2)^3} \{
\tilde{F}\cdot\sigma, \Slash{x}\} \bigl[
(4e_dm_d\rho_d-e_um_u\rho_u)\chi \bar{\theta} \nonumber \\
&+ \chi \{(e_um_d\rho_d-e_dm_u\rho_u)\theta_Q+(
 e_dm_u\rho_d -e_um_d\rho_u)\theta_G\} \nonumber \\
&+(6d_d-d_u) +(\kappa -\frac{1}{2}\xi )
(6e_d\tilde{d}_d -e_u\tilde{d}_u)
\bigr]~.
\end{align}
Thus, $\Pi^{(1)}(x)$ 
is given as
\begin{align}
 \Pi^{(1)}(x) =&\frac{i}{16\pi^4}\langle\bar{q}q\rangle\frac{1}{(x^2)^3}
\{\tilde{F}\cdot\sigma, \Slash{x}\}\bigl[
(1+\beta)^2(4e_dm_d\rho_d-e_um_u\rho_u)\chi \bar{\theta}\nonumber \\
&+(1-\beta^2)\chi\{
(e_dm_u\rho_u-e_um_d\rho_d)\theta_Q+(e_um_d\rho_u-e_dm_u\rho_d)\theta_G
\} \nonumber \\
&+2(3+2\beta+3\beta^2)d_d-(1+\beta)^2d_u \nonumber \\
&+\bigl(\kappa-\frac{1}{2}\xi \bigr)
\{
2(3+2\beta+3\beta^2)e_d\tilde{d}_d-(1+\beta)^2e_u\tilde{d}_u
\}
\bigr]~,
\end{align}
and for $\beta=+1$, the above expression reduces to
\begin{align}
\Pi^{(1)}(x) =&\frac{i}{4\pi^4}\langle \bar{q}q\rangle \frac{1}{(x^2)^3} \{
\tilde{F}\cdot\sigma, \Slash{x}\} \bigl[
(4e_dm_d\rho_d-e_um_u\rho_u)\chi \bar{\theta} \nonumber \\
&+(4d_d-d_u) +(\kappa -\frac{1}{2}\xi )
(4e_d\tilde{d}_d -e_u\tilde{d}_u)
\bigr]~.
\end{align}

Next, we evaluate $\Pi^{(2)}$.
Here, we again use the expressions in Eq.~(\ref{corr_prop}), while
one of the propagators in each correlator is $S^q(x)|_{\rm 1~photon}$ in
this case. The result is given as
\begin{align}
 \Pi^{(2)}_{11}(x) =&\frac{i}{8\pi^4}(2d_d-d_u)\langle \bar{q}q \rangle 
\frac{1}{(x^2)^3} \{\tilde{F}\cdot\sigma, \Slash{x}\}~, \nonumber \\
 \Pi^{(2)}_{12}(x) =& \Pi^{(2)}_{21}(x)=
-\frac{i}{8\pi^4}d_d\langle \bar{q}q \rangle 
\frac{1}{(x^2)^3} \{\tilde{F}\cdot\sigma, \Slash{x}\}~, \nonumber \\
 \Pi^{(2)}_{22}(x) =& 
\frac{i}{8\pi^4}d_u\langle \bar{q}q \rangle 
\frac{1}{(x^2)^3} \{\tilde{F}\cdot\sigma, \Slash{x}\}~,
\end{align}
and they lead to
\begin{align}
  \Pi^{(2)}(x) = &\frac{i}{8\pi^4}\bigl[
2(1-\beta)d_d-(1-\beta^2)d_u
\bigr]\langle\bar{q}q\rangle\frac{1}{(x^2)^3}
\{\tilde{F}\cdot\sigma, \Slash{x}\}~.
\end{align}
Therefore, we find that $\Pi^{(2)}$ vanishes when we take $\beta=+1$:
\begin{equation}
  \Pi^{(2)}(x) = 0~.
\end{equation}

Finally, we study $\Pi^{(3)}$. In this case, we use the equations in
Eq.~(\ref{correlator_propagator}). By using Eq.~(\ref{chichig}),
we find the resultant expressions as
\begin{align}
 \Pi^{(3)}_{11}(x) =&\frac{i}{16\pi^4}\biggl[
-e_d\tilde{d}_d\bigl(\kappa-\frac{1}{2}\xi \bigr)
+(e_d\tilde{d}_u-e_u\tilde{d}_d)(\kappa+\frac{1}{2}\xi \bigr)
\biggr] \langle \bar{q}q \rangle 
\frac{1}{(x^2)^3} \{\tilde{F}\cdot\sigma, \Slash{x}\}~, \nonumber \\
 \Pi^{(3)}_{12}(x) =&\Pi^{(3)}_{21}(x)=
\frac{i}{16\pi^4}
e_d\tilde{d}_d\bigl(\kappa-\frac{1}{2}\xi \bigr)
 \langle \bar{q}q \rangle 
\frac{1}{(x^2)^3} \{\tilde{F}\cdot\sigma, \Slash{x}\}~, \nonumber \\
 \Pi^{(3)}_{22}(x) =&\frac{i}{16\pi^4}\biggl[
-e_d\tilde{d}_d\bigl(\kappa-\frac{1}{2}\xi \bigr)
-(e_d\tilde{d}_u-e_u\tilde{d}_d)(\kappa+\frac{1}{2}\xi \bigr)
\biggr] \langle \bar{q}q \rangle 
\frac{1}{(x^2)^3} \{\tilde{F}\cdot\sigma, \Slash{x}\}~.
\end{align}
Thus, $ \Pi^{(3)}(x)$ is given as
\begin{align}
  \Pi^{(3)}(x) =&\frac{i}{16\pi^4}\langle\bar{q}q\rangle\frac{1}{(x^2)^3}
\{\tilde{F}\cdot\sigma, \Slash{x}\} \nonumber \\
&\times\bigl[
-(1-\beta)^2\bigl(\kappa-\frac{1}{2}\xi\bigr)e_d\tilde{d}_d+
(1-\beta^2)\bigl(\kappa+\frac{1}{2}\xi\bigr)(e_d\tilde{d}_u- e_u \tilde{d}_d)
\bigr]~.
\end{align}
Again, the correlator turns out to vanish for $\beta=+1$:
\begin{equation}
 \Pi^{(3)}(x)=0~.
\end{equation}

Taking the above discussion into account, we conclude that the
correlator $\Pi(x)$ is given as
\begin{align}
 \Pi(x)=&\frac{i}{16\pi^4}\langle\bar{q}q\rangle\frac{1}{(x^2)^3}
\{\tilde{F}\cdot\sigma, \Slash{x}\} \nonumber \\
&\times\bigl[
(1+\beta)^2(4e_dm_d\rho_d-e_um_u\rho_u)\chi \bar{\theta}\nonumber \\
&+(1-\beta^2)\chi\{
(e_dm_u\rho_u-e_um_d\rho_d)\theta_Q+(e_um_d\rho_u-e_dm_u\rho_d)\theta_G
\} \nonumber \\
&+(10+6\beta^2)d_d-(3+2\beta-\beta^2)d_u \nonumber \\
&+\tilde{d}_d\{
2\bigl[(3+2\beta+3\beta^2)
-\frac{1}{2}(1-\beta)^2
\bigr]e_d\bigl(\kappa-\frac{1}{2}\xi\bigr)
-(1-\beta^2)e_u\bigl(\kappa+\frac{1}{2}\xi\bigr)
\} \nonumber \\
&+\tilde{d}_u \{
(1-\beta^2)e_d \bigl(\kappa+\frac{1}{2}\xi\bigr)
-(1+\beta)^2e_u \bigl(\kappa-\frac{1}{2}\xi\bigr)
\}
\bigr]~,
\end{align}
and its Fourier transform is
\begin{align}
 \Pi(q)=&i\int d^4xe^{iq\cdot x} \Pi(x) \nonumber \\
 =&\frac{1}{64\pi^2}\langle\bar{q}q\rangle\log\biggl(
\frac{-q^2}{\Lambda^2}
\biggr)
\{\tilde{F}\cdot\sigma, \Slash{q}\} \nonumber \\
&\times\bigl[
(1+\beta)^2(4e_dm_d\rho_d-e_um_u\rho_u)\chi \bar{\theta}\nonumber \\
&+(1-\beta^2)\chi\{
(e_dm_u\rho_u-e_um_d\rho_d)\theta_Q+(e_um_d\rho_u-e_dm_u\rho_d)\theta_G
\} \nonumber \\
&+(10+6\beta^2)d_d-(3+2\beta-\beta^2)d_u \nonumber \\
&+\tilde{d}_d\{
2\bigl[(3+2\beta+3\beta^2)
-\frac{1}{2}(1-\beta)^2
\bigr]e_d\bigl(\kappa-\frac{1}{2}\xi\bigr)
-(1-\beta^2)e_u\bigl(\kappa+\frac{1}{2}\xi\bigr)
\} \nonumber \\
&+\tilde{d}_u \{
(1-\beta^2)e_d \bigl(\kappa+\frac{1}{2}\xi\bigr)
-(1+\beta)^2e_u \bigl(\kappa-\frac{1}{2}\xi\bigr)
\}
\bigr]~.
\label{sumsum}
\end{align}
Here, a certain ultraviolet mass scale $\Lambda$ is introduced, though
it is irrelevant to our final result. When one sets $\beta=+1$, this
expression reduces to
\begin{align}
\Pi(q)^{\rm (OPE)}
=& \frac{1}{16\pi^2}\langle \bar{q}q\rangle \log\biggl(
\frac{-q^2}{\Lambda^2}
\biggr) \{
\tilde{F}\cdot\sigma, \Slash{q}\} \bigl[
(4e_dm_d\rho_d-e_um_u\rho_u)\chi \bar{\theta} \nonumber \\
&+(4d_d-d_u) +(\kappa -\frac{1}{2}\xi )
(4e_d\tilde{d}_d -e_u\tilde{d}_u)
\bigr]~.
\label{corr_ope}
\end{align}

Equation~(\ref{sumsum}) is corresponding to Eqs.~(9-12) in
Ref.~\cite{hep-ph/0010037}.
After taking $\beta=+1$, we find that the CEDM contribution, {\it
i.e.}, the last term in Eq.~(\ref{corr_ope}) differs from those in the
reference by a factor of 4. In addition, the sign in front of $\xi$ is
opposite to the one in Ref.~\cite{hep-ph/0010037}.

\subsection{Next-to-leading order}

Figures \ref{leading2} and \ref{leading3} yield the next-to-leading
order (NLO) contributions. By using the propagator given in
Eq.~(\ref{photon_propagator}), we evaluate the contribution by the
diagram in Fig.~\ref{leading2} as
\begin{eqnarray}
 \Pi^{(2)}_{11}(x)_{\rm NLO} &=&\frac{i}{32\pi^4}
\langle\bar{q}q\rangle\frac{1}{(x^2)^2}\log(-\Lambda_{IR}^2x^2)
\{\tilde{F}\cdot\sigma, \Slash{x}\} \nonumber \\
&&\times[e_dm_d\rho_d\bar{\theta}+(e_um_u\rho_d-e_dm_d\rho_u)\theta_G
+(e_dm_d\rho_d-e_um_u\rho_u)\theta_Q]~, \nonumber \\
 \Pi^{(2)}_{12}(x)_{\rm NLO} &=& \Pi^{(2)}_{21}(x)_{\rm NLO}\nonumber \\
&=&-\frac{i}{32\pi^4}
\langle\bar{q}q\rangle\frac{1}{(x^2)^2}\log(-\Lambda_{IR}^2x^2)
\{\tilde{F}\cdot\sigma, \Slash{x}\}e_dm_d\rho_d\bar{\theta}~, \nonumber \\
 \Pi^{(2)}_{22}(x)_{\rm NLO} &=&\frac{i}{32\pi^4}
\langle\bar{q}q\rangle\frac{1}{(x^2)^2}\log(-\Lambda_{IR}^2x^2)
\{\tilde{F}\cdot\sigma, \Slash{x}\}
\nonumber \\
&&\times[e_dm_d\rho_d\bar{\theta}-(e_um_u\rho_d-e_dm_d\rho_u)\theta_G
-(e_dm_d\rho_d-e_um_u\rho_u)\theta_Q]~,
\end{eqnarray}
and therefore, $\Pi^{(2)}(x)_{\rm NLO}$ is found to be
\begin{align}
  \Pi^{(2)}(x)_{\rm NLO}=&\frac{i}{32\pi^4}
\langle\bar{q}q\rangle\frac{1}{(x^2)^2}\log(-\Lambda_{IR}^2x^2)
\{\tilde{F}\cdot\sigma, \Slash{x}\} [
(1-\beta)^2e_dm_d\rho_d\bar{\theta} \nonumber \\
&+(1-\beta^2)\{
(e_um_u\rho_d-e_dm_d\rho_u)\theta_G
+(e_dm_d\rho_d-e_um_u\rho_u)\theta_Q
\}
]~.
\label{nlo2}
\end{align}

The gluon contribution illustrated in Fig.~\ref{leading3} is also
calculated by using the propagator displayed in
Eq.~(\ref{gluon_propagator}). The resultant expressions are
\begin{eqnarray}
  \Pi^{(3)}_{11}(x)_{\rm NLO} &=&-\frac{i}{64\pi^4}
\langle\bar{q}q\rangle\frac{1}{(x^2)^2}\log(-\Lambda_{IR}^2x^2)
\{\tilde{F}\cdot\sigma, \Slash{x}\} \nonumber \\
&&\times[e_dm_d\rho_d\bigl(
\kappa+\frac{1}{2}\xi
\bigr)\bar{\theta}+(e_dm_u\rho_d-e_um_d\rho_u)\bigl(
\kappa-\frac{1}{2}\xi
\bigr)\theta_G \nonumber \\
&&+(e_um_d\rho_d-e_dm_u\rho_u)\bigl(
\kappa-\frac{1}{2}\xi
\bigr)\theta_Q]~, \nonumber \\
 \Pi^{(3)}_{12}(x)_{\rm NLO} &=& \Pi^{(3)}_{21}(x)_{\rm NLO}\nonumber \\
&=&\frac{i}{64\pi^4}
\langle\bar{q}q\rangle\frac{1}{(x^2)^2}\log(-\Lambda_{IR}^2x^2)
\{\tilde{F}\cdot\sigma, \Slash{x}\} \bar{\theta}e_dm_d\rho_d\bigl(
\kappa+\frac{1}{2}\xi
\bigr)~, \nonumber \\
  \Pi^{(3)}_{22}(x)_{\rm NLO} &=&-\frac{i}{64\pi^4}
\langle\bar{q}q\rangle\frac{1}{(x^2)^2}\log(-\Lambda_{IR}^2x^2)
\{\tilde{F}\cdot\sigma, \Slash{x}\} \nonumber \\
&&\times[e_dm_d\rho_d\bigl(
\kappa+\frac{1}{2}\xi
\bigr)\bar{\theta}-(e_dm_u\rho_d-e_um_d\rho_u)\bigl(
\kappa-\frac{1}{2}\xi
\bigr)\theta_G \nonumber \\
&&-(e_um_d\rho_d-e_dm_u\rho_u)\bigl(
\kappa-\frac{1}{2}\xi
\bigr)\theta_Q]~, \nonumber \\
\end{eqnarray}
and then, 
\begin{align}
  \Pi^{(3)}(x)_{\rm NLO} =&-\frac{i}{64\pi^4}
\langle\bar{q}q\rangle\frac{1}{(x^2)^2}\log(-\Lambda_{IR}^2x^2)
\{\tilde{F}\cdot\sigma, \Slash{x}\} 
[(1-\beta)^2\bar{\theta}e_dm_d\rho_d\bigl(
\kappa+\frac{1}{2}\xi
\bigr)\nonumber \\
&+(1-\beta^2)\{(e_dm_u\rho_d-e_um_d\rho_u)\theta_G 
+(e_um_d\rho_d-e_dm_u\rho_u)\theta_Q\}\bigl(
\kappa-\frac{1}{2}\xi
\bigr)]~. \nonumber \\
\label{nlo3}
\end{align}
From the results in Eqs.~(\ref{nlo2}) and (\ref{nlo3}), it is found that
taking $\beta=+1$ makes the NLO contributions vanish, as mentioned
before. Thus, we find that the correlator given in Eq.~(\ref{corr_ope})
is valid up to the next-to-leading order.

\section{QCD sum rules}
\label{sec:QCD_Sum_Rules}

In order to derive the QCD sum rules for the present case, we first
extract the coefficient functions of $ \{\tilde{F}\cdot\sigma, \Slash{q}\}$
from both the phenomenological and the OPE correlators, $\Pi^{\rm
(phen)}$ in Eq.~(\ref{corr_phen}) and $\Pi^{\rm (OPE)}$ in
Eq.~(\ref{corr_ope}), respectively:
\begin{align}
 C^{\rm (phen)}(Q^2) &\equiv \frac{1}{2}\biggl[
\frac{\lambda_n^2d_nm_n}{(Q^2+m_n^2)^2}-\frac{A}{Q^2+m_n^2}
\biggr]~,\label{c_phen} \\
 C^{\rm (OPE)}(Q^2) &\equiv\frac{1}{16\pi^2}\langle \bar{q}q\rangle
\Theta
 \log\biggl( \frac{Q^2}{\Lambda^2}
\biggr)~,
\end{align}
with $Q^2\equiv -q^2$ and
\begin{equation}
 \Theta \equiv (4e_dm_d\rho_d-e_um_u\rho_u)\chi \bar{\theta}
+(4d_d-d_u) +(\kappa -\frac{1}{2}\xi )
(4e_d\tilde{d}_d -e_u\tilde{d}_u)~.
\end{equation}
In Eq.~(\ref{c_phen}), we neglect the continuum contribution and think
of $A$ as a constant, as discussed above. The QCD sum rules are obtained
by equating the coefficient functions after the Borel transformation,
{\it i.e.}, 
\begin{equation}
 {\cal B}\bigl[ C^{\rm (phen)}(Q^2)\bigr]=
 {\cal B}\bigl[ C^{\rm (OPE)}(Q^2)\bigr]~,
\end{equation} 
where the Borel transformation of the function $f(Q^2)$ is defined as
\begin{equation}
 {\cal B}\bigl[f(Q^2)\bigr]\equiv \lim_{Q^2,n\to \infty \atop Q^2/n=M^2}
\frac{(Q^2)^{n+1}}{n!}\biggl( \frac{-d}{dQ^2}
\biggr)^n f(Q^2)~,
\end{equation}
with $M$ so-called the Borel mass.
Then, we finally derive the sum rules as follows:
\begin{equation}
\lambda_n^2d_nm_n-AM^2 
=-\Theta \langle \bar{q}q\rangle\frac{M^4}{8\pi^2}e^{\frac{m_n^2}{M^2}}
~.
\label{sum_rule}
\end{equation}
All we have to do is now reduced to determining the Borel mass $M$ and the
coupling $\lambda_n$, as well as estimating the parameter $A$.

\begin{figure}[t]
\begin{center}
\includegraphics[height=70mm]{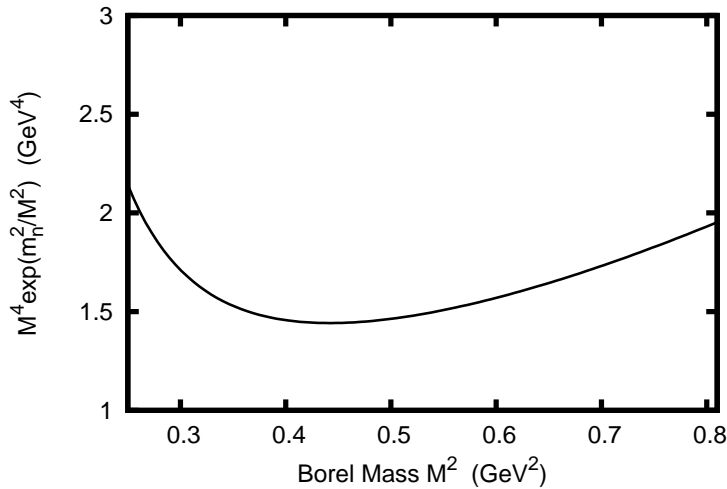}
\caption{Dependence of our sum rules on Borel mass $M^2$. Range of $M$ is
 set to be $0.5~{\rm GeV} \leq M\leq 0.9~{\rm GeV}$.}  
\label{func}
\end{center}
\end{figure}

To illustrate the dependence of the sum rules on the Borel mass, we plot
$M^4e^{\frac{m_n^2}{M^2}}$, which is included in the right-hand side of
Eq.~(\ref{sum_rule}), as a function of the Borel mass squared $M^2$ in
Fig.~\ref{func}. Here, the range of $M$ is set to be $0.5~{\rm GeV} \leq
M\leq 0.9~{\rm GeV}$. From the figure we find that the Borel mass dependence
of our sum rule is moderate in the range of $0.6~{\rm GeV}\lesssim M \lesssim
0.8~{\rm GeV}$. 

\section{Determination of $\lambda_n$ from lattice}
\label{sec:lambda}

The low-energy constant $\lambda_n$ determines the normalization of
the QCD sum rules so that the uncertainties are directly linked to the
final result. We extract its numerical value from the lattice QCD
calculation presented in Ref.~\cite{Aoki:2008ku}, in which the QCD
matrix elements for the proton decay rate are evaluated. In fact, they
evaluate a similar quantity for proton, though the isospin symmetry
allows us to interpret it for the present purpose.

First, we introduce a generic notation for three-quark operators with
an arbitrary spin structure:
\begin{equation}
 {\cal O}^{\Gamma\Gamma^\prime}(\bm{x},t)\equiv
  \epsilon_{abc}\bigl[d^T_a(\bm{x},t)C\Gamma u_b(\bm{x},t)\bigr]
  \Gamma^\prime d_c(\bm{x},t)~,
\end{equation}
with $\Gamma$ and $\Gamma^\prime$ arbitrary $4\times 4$ matrices. In the
current case, the relevant matrices are $R= P_R\equiv
\frac{1}{2}(1+\gamma_5)$ or $L= P_L\equiv \frac{1}{2}(1-\gamma_5)$.
Now we define parameters $\alpha_1$ and $\alpha_2$ as follows:
\begin{align}
 \langle0|{\cal O}^{RL}|N(\bm{p},s)\rangle =&\alpha_1 P_L u_n(\bm{p},s)~,
\nonumber \\
 \langle0|{\cal O}^{LL}|N(\bm{p},s)\rangle =&\alpha_2 P_L u_n(\bm{p},s)~.
\end{align}
The phase definition is fixed such that $\alpha_1$ and $\alpha_2$ are
both real and $\alpha_1<0$. The parity transformation of the above
equations implies that
\begin{align}
 \langle0|{\cal O}^{LR}|N(\bm{p},s)\rangle =&-\alpha_1 P_R u_n(\bm{p},s)~,
\nonumber \\
 \langle0|{\cal O}^{RR}|N(\bm{p},s)\rangle =&-\alpha_2 P_R u_n(\bm{p},s)~.
\end{align}
The interpolating fields $j_1$ and $j_2$ are expressed in terms of the
operators as
\begin{align}
  j_1(x)=&2\bigl({\cal O}^{RL}(x) +{\cal O}^{RR}(x) -{\cal O}^{LL}(x)
  -{\cal O}^{LR}(x)  \bigr)~, \\
  j_2(x)=&2\bigl({\cal O}^{LR}(x) -{\cal O}^{LL}(x) +{\cal O}^{RR}(x)
  -{\cal O}^{RL}(x)  \bigr)~.
\end{align}
Thus their matrix elements between the vacuum and one-particle states
are given as
\begin{align}
 \langle 0| j_1 |N(\bm{p},s)\rangle=&2(\alpha_1-\alpha_2)u_n(\bm{p},s)~, \\
 \langle 0| j_2 |N(\bm{p},s)\rangle=&-2(\alpha_1+\alpha_2)u_n(\bm{p},s)~,
\end{align}
and they lead to
\begin{equation}
 \langle 0|\eta_n|N(\bm{p},s)\rangle =2\bigl[(\alpha_1-\alpha_2)- \beta
  (\alpha_1+ \alpha_2)\bigr]u_n(\bm{p},s)~.
\end{equation}
From the equation we may relate $\lambda_n$ with the parameters
$\alpha_1$ and $\alpha_2$:
\begin{equation}
 \lambda_n(\mu)=2\bigl[(\alpha_1-\alpha_2)- \beta
  (\alpha_1+ \alpha_2)\bigr]~,
\label{lambda_gen_lat}
\end{equation}
with $\mu$ the renormalization scale.
The parameters $\alpha_1$ and $\alpha_2$ at
$\mu = 2$ GeV are estimated in Ref.~\cite{Aoki:2008ku} as
\begin{equation}
 \alpha_1=-0.0112\pm 0.0012_{(\rm stat)} \pm0.0022_{(\rm syst)}~{\rm GeV}^3~,
\end{equation}
\begin{equation}
 \alpha_2=0.0120\pm 0.0013_{(\rm stat)} \pm0.0023_{(\rm syst)}~{\rm GeV}^3~.
\end{equation}
For $\beta=1$, $\lambda_n(\mu=2~{\rm GeV})$ is given as
\begin{align}
 \lambda_n=&-4\alpha_2 \nonumber \\
 =&-0.0480\pm 0.0052_{(\rm stat)} \pm0.0092_{(\rm syst)}~{\rm GeV}^3~.
\end{align}

Since the QCD parameters used here is evaluated at $\mu=1$
GeV, we need to translate the above value of $\lambda_n$ into that of
$\mu=1$ GeV. The one-loop correction for $\lambda_n$ is
\begin{equation}
 \lambda_n(\mu=1~{\rm GeV})=
\biggl(\frac{\alpha_s(1~{\rm GeV})}{\alpha_s(m_c)}\biggr)
^{-\frac{2}{9}}
\biggl(\frac{\alpha_s(m_c)}{\alpha_s(2~{\rm
GeV})}\biggr)
^{-\frac{6}{25}}
\lambda_n(\mu=2~{\rm GeV})~,
\end{equation}
which results in a reduction factor of $\simeq 0.91$. 
As a result, we obtain 
\begin{equation}
 \lambda_n =-0.0436
\pm 0.0047_{(\rm stat)} \pm0.0084_{(\rm syst)}~{\rm GeV}^3~,
\label{lattice_lambda_1}
\end{equation}
for $\beta=+1$.

Let us compare the value of $\lambda_n$ obtained here with those used in
the previous works. In Ref.~\cite{hep-ph/9904483}, for
example, they exploit the values for $\lambda_n$ evaluated in
Ref.~\cite{arXiv:nucl-th/9510051}\footnote{
Note that the notation used in Ref.~\cite{arXiv:nucl-th/9510051} is
different from ours:
\begin{equation}
 \lambda_n=2\lambda_{\cal O}=\frac{2}{(2\pi)^2}\tilde{\lambda}_{\cal O}~.
\end{equation}
Also, notice that there is some difference between the results described in
Ref.~\cite{arXiv:nucl-th/9510051} and the corresponding expressions
shown in Ref.~\cite{hep-ph/9904483}.
} by using the QCD sum rules. Two
Dirac-$\gamma$ structures, $\1$ and $\Slash{p}$, provide different sum
rules.
 As
evaluated in Ref.~\cite{arXiv:nucl-th/9510051}, these two sum rules
yield relatively small values for $\lambda_n$; the lattice QCD value is
several times larger than the values evaluated by using the QCD sum
rules. The author in Ref.~\cite{arXiv:nucl-th/9510051} also estimates
the error for these values. It is about $30$ \% for the sum rules
result, while $20$ \% for the lattice QCD result. The lattice QCD result
might have a uncertainty in the chiral extrapolation. Since there is no
more guiding principle for judging  which estimation is valid, we
exploit the lattice QCD result in Eq.~(\ref{lattice_lambda_1}) because
this choice leads to rather conservative constraint for CP-violating
sources.

\section{Results}
\label{sec:result}

Now we estimate the neutron EDM by using the results obtained
above. First of all, we rewrite the sum rules in Eq.~(\ref{sum_rule}) in
a simple form:
\begin{equation}
 c_0 + c_1 x = f(x)~,
\label{simple}
\end{equation}
where $x=M^2$ and
\begin{equation}
 f(x)\equiv \frac{x^2}{8\pi^2}\exp\bigl(\frac{m_n^2}{x}\bigr)~,
~~~~~c_0\equiv \frac{d_n \lambda_n^2 m_n}{-\Theta\langle \bar{q} q
\rangle }~,~~~~~
c_1\equiv \frac{-A}{-\Theta\langle \bar{q} q
\rangle}~.
\end{equation}
The right-hand side of Eq.~(\ref{simple}) describes the behavior of the
coefficient function obtained from the OPE calculation, while the
left-hand side represents the phenomenological one. The first
and second terms in the left-hand side correspond to the double and
single pole contributions, {\it i.e.}, the first and second terms in
Eq.~(\ref{phenomenology_f}), respectively. Once given a
Borel mass point $x=M^2$, one may readily pick out $c_0$ and $c_1$ from
the tangent line to the function $f(x)$ at the point. Then, they are
expressed as the functions of $x$ as 
\begin{align}
 c_0(x) =& \frac{1}{8\pi^2}(m_n^2x-x^2)\exp\bigl(\frac{m_n^2}{x} \bigr)~,
\nonumber \\
 c_1(x) =& \frac{1}{8\pi^2}(2x-m_n^2)\exp\bigl(\frac{m_n^2}{x} \bigr)~.
\end{align}
From these expressions, it is found that the single pole contribution
vanishes at $x=m_n^2/2$. Since the parameter $A$ is unknown, this choice
of $x$ is favorable in order to estimate the double pole
contribution. Then, at this point the value of $c_0$ is
\begin{equation}
 c_0=1.8\times 10^{-2}~~~~~~~({\rm for}~x=m_n^2/2)~,
\end{equation} 
and therefore, the neutron EDM $d_n$ is evaluated as
\begin{align}
 d_n=&\frac{-c_0\langle \bar{q} q\rangle}{\lambda_n^2m_n}\Theta~
=1.2\times 10^{-1} ~\Theta~.
\label{result0}
\end{align}
Here we take $\langle \bar{q} q\rangle=-(0.225~{\rm GeV})^3$~. 

\begin{figure}[t]
\begin{center}
\includegraphics[height=70mm]{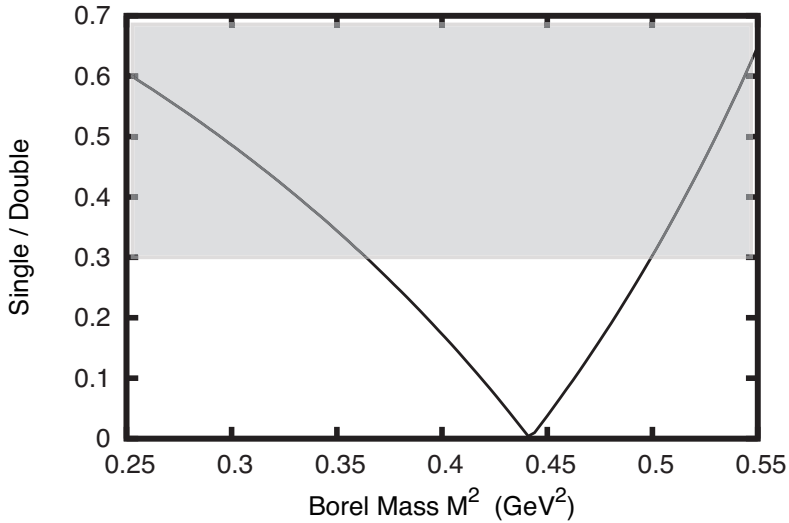}
\caption{Ratio of single and double pole contributions as a function of
 Borel mass $M^2$. Shaded region illustrates that single pole
 contribution is more than 30\% of double pole contribution.}  
\label{error}
\end{center}
\end{figure}

The
choice of the Borel mass, $M^2=m_n^2/2$, is, however, quite arbitrary,
and deviation from the above result due to the different choice of the
Borel mass should be taken into account as theoretical
uncertainty. In Fig.~\ref{error}, we plot the ratio of the single and
double pole contributions as a function of $M^2$. From this figure we
find that the single pole contribution rapidly increases when the Borel
mass is varied from $M^2=m_n^2/2$. We here assume our sum rules to be
valid within the region of the Borel mass in which the single pole
contribution is less than $30~\%$ of the double pole contribution. This
assumption leads to $0.36~{\rm GeV}^2<M^2<0.50$~GeV$^2$, and in this
region, $d_n$ takes the following values: 
\begin{equation}
 d_n=1.2~^{+0.6}_{-0.3}\times 10^{-1}~\Theta~.
\label{result1}
\end{equation}
Here, the lower value corresponds to the upper limit of the Borel mass,
and vice versa. 

Next, we discuss the uncertainty of the OPE calculation. In this
case, the truncation of the OPE leads to the uncertainty. Let us
estimate it by evaluating the relative size of the higher-order
contributions. Among them, the four-quark condensates such as $\langle
\bar{q}q \bar{q}q\rangle$ are expected to yield sizable contributions,
since they are free from loop suppression. On the assumption that these
contributions vanish when one takes the quark masses to be zero, we
expect that they are suppressed at least a factor of $\langle \bar{q}q
\rangle ^{\frac{2}{3}}/M^2\simeq 0.1$. Therefore, the uncertainty of the
OPE calculation is estimated to be ${\cal O}(10)~\%$. 

Taking the above discussion into account, we finally evaluate the
neutron EDM with theoretical uncertainty as follows:
\begin{equation}
 d_n=1.2~^{+0.6}_{-0.3}\pm 0.1~^{+0.7}_{-0.4}\times 10^{-1} ~\Theta~,
\end{equation}
where the first uncertainty stems from the phenomenological
calculation while the second one comes from the approximation in the
OPE. We also include uncertainties originate from those in $\lambda_n$
(See Eq.~(\ref{lattice_lambda_1}).), which is indicated by the third
error in the above equation\footnote{
We approximate the error in $\lambda_n$ as the r.m.s. of the
statistical and systematic errors displayed in Eq.~(\ref{lattice_lambda_1}).
}. After all,
it is found that there is almost
${\cal O}(1)$ factor of uncertainty in our sum rule
calculation.

Let us compare the result with those obtained by using the values of
  $\lambda_n$ calculated with the QCD sum rules. In
  Ref.~\cite{Ioffe:1983ju}, the authors adopt $\lambda_n^2 =
  1.05/(2\pi)^4$~GeV$^3$ for $\beta=-1$ from the QCD sum rules derived
  for the nucleon mass. The $\lambda_n$ for $\beta=-1$ is equal to that
  for $\beta=1$ in the non-relativistic quark limit. In
  Ref.~\cite{Pospelov:2005pr}, it is shown that the neutron EDM
  prediction in the non-relativistic quark model with the SU(6) spin-flavor
  symmetry is derived in the QCD sum rules 
  using the value of $\lambda_n$.  By substituting the value into
  Eq.~(\ref{result0}), one obtains $d_n=3.3\times 10^{-1}~\Theta$. The
  ``realistic'' value of $\lambda_n$ evaluated by the QCD sum rules in
  Ref.~\cite{arXiv:nucl-th/9510051}, $\lambda_n \simeq 0.022$~GeV$^3$,
  leads to a slightly larger result: $d_n=4.6\times
  10^{-1}~\Theta$. Thus, the overall factors of $d_n$ of these results
  are several times larger than that of our result.

For one's convenience, we substitute the numerical values for the
QCD parameters in Eq.~(\ref{result1}). Here we take $m_0^2=0.8~{\rm
GeV}^2$ , $\chi=-5.7\pm0.6~{\rm GeV}^{-2}$, $\xi=-0.74\pm 0.2$, and
$\kappa= -0.34\pm 0.1$ \cite{Belyaev:1982sa, Kogan:1991en}. Then, 
with those parameters the center values, we find
\begin{align}
 d_n =4.2\times 10^{-17}\bar{\theta}~[e~{\rm cm}]+0.47d_d -0.12d_u +
e(-0.18\tilde{d}_u+0.18\tilde{d}_d-0.008\tilde{d}_s)~.
\end{align}
The contributions from $\bar{\theta}$ and
  the quark CEDMs to $d_n$ may be changed furthermore by about $\pm
  10\%$, mainly 
  due to the theoretical uncertainty of $\chi$.

\section {Under the Peccei-Quinn symmetry}
\label{sec:pq}
%
It is known that $O(1)$ $\bar{\theta}$ induces too large neutron EDM,
the strong CP problem. The Peccei-Quinn (PQ) symmetry is
one of the solutions for the strong CP problem.
If  the PQ symmetry is introduced, $\bar{\theta}$ vanishes 
dynamically. However, if the quark CEDMs are non-vanishing, a linear term
is induced to the axion potential \cite{Bigi:1991rh},
\begin{equation}
 V(\bar{\theta})=K' \bar{\theta}-\frac{1}{2} K\bar{\theta}^2~,
\end{equation}
where $K$ is the topological susceptibility
\begin{equation}
K=-i \lim_{k\to 0} i \int d^4x e^{ikx}
\langle T\left[ \frac{\alpha_s}{8\pi}G\tilde{G}(x)
\frac{\alpha_s}{8\pi}G\tilde{G}(0)\right]
\rangle~,
\end{equation}
and $K'$ is calculated by
\begin{equation}
K'=-i \lim_{k\to 0} i \int d^4x e^{ikx}
\langle T [\frac{\alpha_s}{8\pi}G\tilde{G}(x)
\frac{i}{2}\sum_{q=u,d,s}\tilde{d}_q\bar{q}g_s(G\cdot\sigma)\gamma_5q(0)
]\rangle~.
\end{equation}
Minimizing the axion potential, an effective $\theta$ term is induced
from the quark CEDM,
\begin{equation}
 \bar{\theta}_\mathrm{ind}=-\frac{K'}{K}=\frac{m_0^2}{2}\sum_{q=u,d,s}
\frac{\tilde{d}_q}{m_q}~.
\end{equation}

Taking the induced $\theta$ term into account, the neutron EDM in the
presence of the PQ symmetry is estimated as
\begin{equation}
 d_n^\mathrm{PQ}=1.2~^{+0.6}_{-0.3}\pm 0.1~^{+0.7}_{-0.4}
\times 10^{-1}~\Theta^\mathrm{PQ}~,
\end{equation}
where
\begin{equation}
\Theta^\mathrm{PQ}= 4d_d-d_u+
\biggl (\frac{m_0^2}{2}\chi +\kappa -\frac{1}{2}\xi \biggr)
(4e_d\tilde{d}_d -e_u\tilde{d}_u)~.
\end{equation}
The contributions from the strange quark CEDM are cancelled in the 
presence of the PQ symmetry \cite{hep-ph/0010037}.
Again, we substitute the numerical values for the QCD parameters as
presented in the previous section. The result is
\begin{equation}
  d_n^\mathrm{PQ}=0.47d_d-0.12d_u+e(0.35\tilde{d}_d+0.17\tilde{d}_u)~.
\end{equation}

\section{Conclusion and discussion}
\label{sec:coclusion}

We have studied the neutron EDM induced by the CP-violating
interactions up to the dimension-five operators. In order to derive
the relation between the CP-violating interactions and the neutron
EDM, we have used the QCD sum rule technique.  There are several
phenomenological parameters to estimate the relation numerically.
Pospelov and Ritz also analysed the neutron EDM using the QCD sum
rules \cite{hep-ph/9904483, hep-ph/0010037} and they determined the
low-energy constant $\lambda_n$ within the framework in the QCD sum
rules.  On the other hand, we have extracted the $\lambda_n$ parameter
from lattice calculations.  This approach allows us to reduce a
theoretical uncertainty and leads to a conservative constraint on the
CP violations.  Our result is about 70 \% smaller compared with the
one obtained by Pospelov and Ritz.  There still remains a sizable
uncertainty resulting from the QCD sum rules itself due to a choice of
the Borel mass scale.  We have estimated the uncertainty from the
Borel mass scale assuming that the single pole contribution is less
than 30 \% of the double pole contributions. This assumption leads to
the theoretical error of about ${\cal O}(1)$.

Finally, we briefly comment on the contribution of the CP-violating
dimension-six operators. Among the dimension-six operators, the 
following operator, called Weinberg operator,
\begin{eqnarray}
{\cal L}\Slash{_{CP}}&=&\frac{1}{3}w
 f_{ABC}G^{A}_{\mu\nu}\tilde{G}^{B\nu\lambda}
G^{C\mu}_\lambda~,
\nonumber
\end{eqnarray}
might be comparable to the quark EDM and CEDM
contributions since they suffer from the chiral suppression.
The other CP-violating dimension-six operators are effective four-quark
operators of light quarks, which are negligible in the neutron EDM in
typical high-energy models since the Wilson coefficients are
suppressed by the quark masses\footnote{If the CP-violating four-quark
  operators include heavy quarks, the CEDMs for the light quarks and
  the Weinberg operators are radiatively induced by integration of the 
heavy quarks as shown in
  Ref.~\cite{Hisano:2012cc}.}.
In our QCD sum rule calculation it is found that the contribution from the Weinberg operator is ${\cal
  O}(\langle \bar{q}q \rangle^2 )$ and thus sub-dominant. There are a
lot of discussions on the significance of the Weinberg operator
\cite{Demir:2002gg}, though no consensus has been reached yet.

\section*{Acknowledgments}
One of the authors (JH) appreciates useful discussion with Yasumichi Aoki.
This work is supported by Grant-in-Aid for Scientific research from
the Ministry of Education, Science, Sports, and Culture (MEXT), Japan,
No. 20244037, No. 20540252, No. 22244021 and No.23104011 (JH), and
also by World Premier International Research Center Initiative (WPI
Initiative), MEXT, Japan. The work of NN is supported by
Research Fellowships of the Japan Society for the Promotion of Science
for Young Scientists.

\newpage

\section*{Appendix}
\appendix

In this Appendix, we list some techniques which we use to carry out the
calculation. 

\section{Quark condensates on the
 CP-violating background}
\label{sec:quark_condensates}

In this section, we discuss the effects of the CP-violating interactions
on the quark condensates as well as on the quark and gluon background
fields. We begin by estimating it for the generic quark bi-linear condensate
$\langle 0|\bar{q}\Gamma q |0\rangle$, with $\Gamma$ a $4\times 4$ 
constant matrix for which the quark bi-linear $\bar{q}\Gamma q$ is an
Hermitian operator. Then, by using the results obtained there, we derive
the relations between the quark condensates and the background fields.

First, we evaluate the quark bi-linear $\bar{q}\Gamma q$ on the
CP-violating background. The contribution of the QCD $\theta$ term at
the leading order is evaluated as \cite{Shifman:1979if} 
\begin{equation}
 \langle 0| \bar{q}\Gamma q |0\rangle_{\theta_G}
=i\int d^4x \langle 0 |T\{
\bar{q}(0)\Gamma q(0)\theta_G \frac{\alpha_s}{8\pi}G^a_{\mu\nu}(x)
\tilde{G}^{a\mu\nu}(x)
\}|0\rangle~+{\cal O}(\theta^2)~.
\end{equation}
Substituting Eq.~(\ref{axial_anomaly}) into the above expression, we
obtain 
\begin{equation}
 \langle 0| \bar{q}\Gamma q |0\rangle_{\theta_G}
=i\int d^4x \langle 0 |T\{
\bar{q}(0)\Gamma q(0)\frac{\theta_G}{2}[\partial^\mu J_{5\mu}(x)-2i
\sum_{q=u,d,s}m_q\rho_q\bar{q}(x)\gamma_5 q(x)
]
\}|0\rangle+{\cal O}(\theta^2)~,
\end{equation}
with $\rho_q=\theta_q/\theta_Q$.
The first term in the equation is calculated with the aid of the
integration by parts:
\begin{eqnarray}
i\int d^4x \langle 0 |T\{
\bar{q}(0)\Gamma q(0)\frac{\theta_G}{2}\partial^\mu J_{5\mu}(x)
\}|0\rangle 
&=&
-\frac{i\theta_G}{2}\int d^4x\langle 0 |[J^0_5(x),\bar{q}(0)\Gamma
 q(0)]\delta (x^0)|0\rangle \nonumber \\ 
&=&\frac{i\theta_G}{2}\rho_q\langle 0| \bar{q}\{\gamma_5,
 \Gamma\}q|0\rangle~. 
\end{eqnarray}
For the second term, we insert the intermediate states and keep only the
contributions of the one-particle states of the pseudo Nambu-Goldstone
bosons $\pi^0$ and $\eta^0$:
\begin{align}
 &-i\sum_{q=u,d,s}\int d^4x \langle 0 |T\{
\bar{q}(0)\Gamma q(0)\theta_G
m_q\rho_q\bar{q}(x)i\gamma_5 q(x)
\}|0\rangle \nonumber \\
=&-\frac{\theta_G}{f_\pi m_\pi^2}(m_u\rho_u-m_d\rho_d)
\langle \bar{q}q\rangle
\langle 0| \bar{q}\Gamma q|\pi^0\rangle \nonumber \\
&-\frac{\theta_G}{\sqrt{3}f_\pi m_\eta^2}(m_u\rho_u+m_d\rho_d-2m_s\rho_s)
\langle \bar{q}q\rangle
\langle 0| \bar{q}\Gamma q|\eta^0\rangle~,
\end{align}
where $f_\pi$ is the pion decay constant\footnote{
We use the PCAC relation for $\pi^0$,
\begin{equation}
 \partial_\mu J_A^\mu(x)=-f_\pi m_\pi^2 \pi(x)~,
\end{equation}
and a similar relation for $\eta^0$.
}, and $m_\pi$ and $m_\eta$ denote the masses of $\pi^0$ and $\eta^0$,
respectively\footnote{
The effect of the $\pi^0$-$\eta^0$ mixing is suppressed by a small
factor of $(m_u-m_d)/m_s$ and we ignore it for brevity.
}.
As a result, we obtain
\begin{align}
  \langle 0| \bar{q}\Gamma q |0\rangle_{\theta_G}=&
\frac{i\theta_G}{2}\rho_q\langle 0| \bar{q}\{\gamma_5,
 \Gamma\}q|0\rangle \nonumber \\
&-\frac{\theta_G}{f_\pi m_\pi^2}(m_u\rho_u-m_d\rho_d)
\langle \bar{q}q\rangle
\langle 0| \bar{q}\Gamma q|\pi^0\rangle \nonumber \\
&-\frac{\theta_G}{\sqrt{3}f_\pi m_\eta^2}(m_u\rho_u+m_d\rho_d-2m_s\rho_s)
\langle \bar{q}q\rangle
\langle 0| \bar{q}\Gamma q|\eta^0\rangle~.
\end{align}

Other contributions also may be evaluated through a similar procedure. 
For the contribution of the $\gamma_5$-mass terms, 
\begin{align}
  \langle 0| \bar{q}\Gamma q |0\rangle_{\theta_q}=&
-\frac{\theta_Q}{f_\pi m_\pi^2}(m_u\rho_u-m_d\rho_d)
\langle \bar{q}q\rangle
\langle 0| \bar{q}\Gamma q|\pi^0\rangle \nonumber \\
&-\frac{\theta_Q}{\sqrt{3}f_\pi m_\eta^2}(m_u\rho_u+m_d\rho_d-2m_s\rho_s)
\langle \bar{q}q\rangle
\langle 0| \bar{q}\Gamma q|\eta^0\rangle~,
\end{align}
while for the contribution of the quark CEDM terms,
\begin{align}
  \langle 0| \bar{q}\Gamma q |0\rangle_{q\rm CEDM}=&
\frac{1}{f_\pi m_\pi^2}\frac{m_0^2}{2}(\tilde{d}_u-\tilde{d}_d)
\langle \bar{q}q\rangle
\langle 0| \bar{q}\Gamma q|\pi^0\rangle \nonumber \\
&+\frac{1}{\sqrt{3}f_\pi
 m_\eta^2}\frac{m_0^2}{2}(\tilde{d}_u+\tilde{d}_d-2\tilde{d}_s) 
\langle \bar{q}q\rangle
\langle 0| \bar{q}\Gamma q|\eta^0\rangle~.
\end{align}
Furthermore, it is found that the quark EDMs induce no contribution.

Taking all of the contributions into account, we obtain the
CP-violating contribution to the quark condensates as
\begin{align}
  \langle 0| \bar{q}\Gamma q |0\rangle_{\tiny \Slash{\rm CP}}=&
\frac{i\theta_G}{2}\rho_q\langle 0| \bar{q}\{\gamma_5,
 \Gamma\}q|0\rangle \nonumber \\
&+\frac{1}{f_\pi m_\pi^2}\biggl[\frac{m_0^2}{2}(\tilde{d}_u-\tilde{d}_d)
-\bar{\theta}(m_u\rho_u-m_d\rho_d)\biggr]
\langle \bar{q}q\rangle
\langle 0| \bar{q}\Gamma q|\pi^0\rangle \nonumber \\
&+\frac{1}{\sqrt{3}f_\pi
 m_\eta^2}\biggl[\frac{m_0^2}{2}(\tilde{d}_u+\tilde{d}_d-2\tilde{d}_s)
 -\bar{\theta}(m_u\rho_u+m_d\rho_d-2m_s\rho_s)\biggr] 
\langle \bar{q}q\rangle
\langle 0| \bar{q}\Gamma q|\eta^0\rangle \nonumber \\
=&
\frac{i\theta_G}{2}\rho_q\langle 0| \bar{q}\{\gamma_5,
 \Gamma\}q|0\rangle~.
\label{app:condensate_CP}
\end{align}
Here, the second equality comes from the conditions in
Eq.~(\ref{tadpole_conditions2}). Thus, the choice of the quark mass
phases in Eq.~(\ref{quark_mass_phases}) reduces the contribution of the
CP-violating interactions to the vacuum condensates into a quite simple
expression. 

Note that the CP-violating contribution to the quark condensates
vanishes in the basis where the $\theta$ term is completely rotated out
into the imaginary mass term. Thus, the choice of this basis, which is
often adopted in the chiral Lagrangian approach, simplifies the
calculation. In our paper, however, we remain in a
general basis in order to display each contribution explicitly.

Next, we discuss the way of translating the quark and gluon
background fields into their condensates. 
To begin with, we consider a single quark line
$\chi^q_{a\alpha}(x)\bar{\chi}^q_{b\beta}(0)$. In this case, it is
related with the quark condensate as follows:
\begin{equation}
 \chi^q_{a\alpha}(x)\bar{\chi}^q_{b\beta}(0)=
\langle \Omega{\Slash{_\mathrm{CP}}}|
q_{a\alpha}(x)\bar{q}_{b\beta}(0) |\Omega{\Slash{_\mathrm{CP}}}
\rangle_F~.
\end{equation}
Using the Fierz identity, the right-hand side of the expression
leads to
\begin{eqnarray}
\langle \Omega{\Slash{_\mathrm{CP}}}|
q_{a\alpha}(x)\bar{q}_{b\beta}(0) |\Omega{\Slash{_\mathrm{CP}}}
\rangle_F
=&-&\frac{\delta_{ab}}{12}
\biggl[
\langle \bar{q}(0)q(x)\rangle_{F,{\tiny \Slash{\rm CP}}}
+\gamma_5 \langle \bar{q}(0)\gamma_5q(x)\rangle_{F,{\tiny \Slash{\rm
CP}}}\nonumber \\
&+&\gamma^\mu \langle \bar{q}(0)\gamma_\mu q(x)\rangle_{F,{\tiny \Slash{\rm
CP}}} 
-\gamma^\mu\gamma_5 \langle \bar{q}(0)\gamma_\mu\gamma_5
 q(x)\rangle_{F,{\tiny \Slash{\rm CP}}} \nonumber \\
&+&\frac{1}{2}\sigma^{\mu\nu} \langle \bar{q}(0)\sigma_{\mu\nu} 
q(x)\rangle_{F,{\tiny \Slash{\rm CP}}}
\biggr]_{\alpha\beta}~.
\end{eqnarray}
These quark condensate terms are evaluated by 
conducting the short-distance expansion of the quark field 
in the Fock-Schwinger gauge as
\begin{equation}
 q(x)=q(0)+x^\mu D_\mu q(0)+\dots~.
\label{quark_field_expansion}
\end{equation}
We note that in this gauge one does not need to care about Wilson-line
operators for the quark fields. (See Appendix~\ref{wilson_line}.)

In the case of CP-even vacuum, the Lorentz and CP invariance of vacuum
tell us that
\begin{equation}
 \langle \bar{q}\gamma_5q\rangle =
\langle \bar{q}\gamma_\mu q\rangle =
\langle \bar{q}\gamma_\mu\gamma_5q\rangle =0~.
\end{equation}
On the other hand, on an electromagnetic background, $\langle
\bar{q}\sigma_{\mu\nu}q\rangle$ may have non-zero VEV proportional to
the electromagnetic field strength $F_{\mu\nu}$. The electromagnetic
field dependence for quark condensates is given as 
\begin{equation}
 \langle \bar{q}\sigma_{\mu\nu}q\rangle_F   =\chi _q
  F_{\mu\nu}\langle \bar{q}q\rangle ,
\end{equation}
where $\chi_q$ is called the quark condensate magnetic
susceptibility \cite{Ioffe:1983ju}. Similar parametrization is used
for the condensates including the gluon background field:
\begin{equation}
 g_s \langle \bar{q}G^A_{\mu\nu}T^Aq\rangle_F=\kappa _q F_{\mu\nu}
\langle \bar{q}q\rangle~,
\end{equation}
\begin{equation}
 2 g_s \langle \bar{q}\gamma_5 \tilde{G}^A_{\mu\nu}T^Aq\rangle_F
= i\xi _q F_{\mu\nu}
\langle \bar{q}q\rangle~.
\end{equation}
As in Ref.~\cite{Ioffe:1983ju}, we assume $\chi_q$, $\kappa_q$ and
$\xi_q$ to be proportional to the quark charge:
\begin{equation}
 \chi_q=e_q \chi, ~~~~~~\kappa_q=e_q \kappa,~~~~~~\xi_q=e_q\xi~.
\end{equation}
This assumption corresponds to neglecting of the closed-loop
contribution with gluon exchange.

Now let us consider the effect of the CP-violating interaction in
Eq.~(\ref{Lagrangian}) to the quark condensates. 
By using Eq.~(\ref{app:condensate_CP}) and the expansion in
Eq.~(\ref{quark_field_expansion}), we evaluate each quark condensate
on the CP-violating background as follows (with omitting the
subscriptions $F$ and $\Slash{\rm CP}$ for simplicity as long as it is
not confusing):
\begin{equation}
 \langle \bar{q}(0)q(x)\rangle  
= \langle \bar{q}q\rangle~,
\end{equation} 
\begin{equation}
 \langle \bar{q}(0)\gamma_5 q(x)\rangle
= \langle \bar{q}\gamma_5q\rangle_{\tiny \Slash{\rm CP}}
=i\theta_G\rho_q\langle\bar{q}q\rangle~,
\end{equation} 
\begin{align}
\langle \bar{q}(0)\gamma_\mu q(x)\rangle 
=&x^\nu \langle \bar{q}\gamma_\mu D_\nu q\rangle
\nonumber \\
=&\frac{1}{2}x^\nu\langle \bar{q}\{
\gamma_\mu D_\nu +\gamma_\nu D_\mu
\} q\rangle+
\frac{1}{2}x^\nu\langle \bar{q}\{
\gamma_\mu D_\nu -\gamma_\nu D_\mu
\} q\rangle \nonumber  \\
=&\frac{1}{4}x^\nu g_{\mu\nu} \langle \bar{q}\Slash{D}q\rangle
+\frac{i}{4}x^\nu \langle
 \bar{q}[\Slash{D},\sigma_{\mu\nu}]q\rangle \nonumber \\
=&-\frac{i}{4}m_q x_\mu \langle\bar{q}q\rangle~,
\end{align}
where we use the classical equations of motion in the quark condensates
and move the covariant derivatives with help of total
derivative. The validity of this procedure is discussed in
Appendix~\ref{eom}.  Furthermore,
\begin{align}
\langle \bar{q}(0)\gamma_\mu \gamma_5 q(x)\rangle 
=&x^\nu \langle \bar{q}\gamma_\mu D_\nu \gamma_5 q\rangle
\nonumber \\
=&\frac{1}{4}x^\nu g_{\mu\nu} \langle \bar{q}\Slash{D}\gamma_5 q\rangle
+\frac{i}{4}x^\nu \langle
 \bar{q}[\Slash{D},\sigma_{\mu\nu}]\gamma_5 q\rangle \nonumber \\
=&\frac{i}{2}m_qe_q\chi(\bar{\theta}\rho_qF_{\mu\nu}+\tilde{F}_{\mu\nu})
x^\nu \langle \bar{q}q\rangle
+\frac{i}{2}(d_q+[\kappa-\frac{1}{2}\xi]e_q\tilde{d}_q)x^\nu F_{\mu\nu}
\langle\bar{q}q\rangle~,
\end{align}
and
\begin{align}
\langle \bar{q}(0)\sigma_{\mu\nu} q(x)\rangle 
 =&\langle \bar{q}\sigma_{\mu\nu} q\rangle 
=\langle \bar{q}\sigma_{\mu\nu} q\rangle_{\rm CP~even} 
+\langle \bar{q}\sigma_{\mu\nu}
 q\rangle_{\tiny \Slash{\rm CP}} \nonumber \\
=&e_q\chi
 [F_{\mu\nu}-\theta_G\rho_q\tilde{F}_{\mu\nu}]\langle\bar{q}q\rangle~. 
\end{align}
Taking the above discussion into account and using the relation,  
\begin{equation}
 F_{\mu\nu}x^\mu\gamma^\nu\gamma_5=+\frac{1}{4}\{
\tilde{F}\cdot \sigma, \Slash{x}
\}~,
\end{equation}
and
\begin{equation}
 F\cdot \sigma=i\tilde{F}\cdot\sigma\gamma_5~,
\end{equation}
we finally obtain the expression for the single quark line as follows:
\begin{align}
  \chi^q_{a\alpha}(x)\bar{\chi}^q_{b\beta}(0)
=&-\frac{\delta_{ab}}{12}\left(
1+i\theta_G\rho_q\gamma_5
\right)_{\alpha\beta}
\langle\bar{q}q\rangle
+\frac{i}{48}\delta_{ab}\Slash{x}_{\alpha\beta}m_q
\langle\bar{q}q\rangle \nonumber \\ 
 &-\frac{i}{96}\delta_{ab}\left[\bar{\theta}m_q\rho_q e_q \chi 
+d_q+(\kappa-\frac{1}{2}\xi)e_q\tilde{d}_q
\right]
\{
\tilde{F}\cdot \sigma, \Slash{x}
\}_{\alpha\beta}
\langle\bar{q}q\rangle \nonumber \\
&+\frac{i}{96}m_qe_q \chi \delta_{ab}
\{
F\cdot \sigma, \Slash{x}
\}_{\alpha\beta}
\langle\bar{q}q\rangle \nonumber \\ 
 &-\frac{i}{24}e_q \chi \delta_{ab}
\left(\tilde{F}\cdot\sigma\gamma_5[1+i
\rho_q\theta_G\gamma_5]
\right)_{\alpha\beta}
\langle\bar{q}q\rangle ~.
\end{align}

Lastly, we evaluate the interaction part of the quark and gluon background
fields, 
\begin{equation}
 g_s\chi^q_{a\alpha}(x) \bar{\chi}^q_{b\beta}(0)
[G_{\mu\nu}]_{cd}=
\langle g_s q_{a\alpha}(x)[G_{\mu\nu}]_{cd} \bar{q}_{b\beta}(0)
\rangle_{F,{\tiny \Slash{\rm
CP}}}~.
\end{equation}
Again, we use the Fierz identity and the short-distance expansion of the
quark field, and through a similar calculation, we find the following
results: 
\begin{align}
  g_s\chi^q_{a\alpha}(x) \bar{\chi}^q_{b\beta}(0)
[G_{\mu\nu}]_{cd} =
&-
 \frac{1}{32}(\delta_{ad}\delta_{bc}-\frac{1}{3}\delta_{ab}\delta_{cd})
\langle\bar{q}q\rangle
\nonumber \\
&\times\biggl[
e_q(\kappa F_{\mu\nu}-\frac{i}{2}\xi \tilde{F}_{\mu\nu}\gamma_5)
(1+i\theta_G \rho_q \gamma_5)
 \nonumber \\
&-\frac{i}{4}m_qe_q\Slash{x}(\kappa F_{\mu\nu}+\frac{1}{2}\bar{\theta}
 \rho_q \xi \tilde{F}_{\mu\nu}) \nonumber \\
&-\frac{i}{24}m_q m_0^2 \epsilon_{\mu\nu\rho\sigma} x^\rho \gamma^\sigma
 \gamma_5 -\frac{i}{24}\bar{\theta}m_q\rho_qm_0^2 (x_\mu \gamma_\nu
 \gamma_5 -x_\nu \gamma_\mu \gamma_5) \nonumber \\
&-\frac{1}{12}m_0^2 \sigma_{\mu\nu}
-\frac{i}{12}m_0^2 \theta_G \rho_q \sigma_{\mu\nu} \gamma_5 
\biggr]_{\alpha\beta}~.
\end{align}

\section{Equations of motion}
\label{eom}

Let us discuss the validity of using classical equations of motion
for quark condensates. We investigate the following quantity:
\begin{equation}
 \langle 0 |\bar{q}\Gamma (i\Slash{D}-m_q)q|0\rangle _{\theta} ~,
\label{fq}
\end{equation}
where $\Gamma$ is a ($c$-number) $4\times 4$ matrix.  The subscript
indicates that this quantity is evaluated in the $\theta$ vacuum. One may
readily generalize the discussion here for the case with other
CP-violating sources. The discussion presented in this section is
based on Ref.~\cite{CALT-68-765}.

First, we define the generating functional $Z[\eta]_{\theta}$ on the
same background: 
\begin{equation}
 Z[\eta]_{\theta}\equiv \int{\cal D}\bar{q}{\cal D}q\exp\left[
i\int d^4x\left\{
{\cal L}+\eta\bar{q}\Gamma (i\Slash{D}-m_q)q
\right\}
\right]~.
\end{equation}
Here, the Lagrangian density ${\cal L}$ is
\begin{equation}
 {\cal L}=\bar{q}(i\Slash{D}-m_q)q~.
\end{equation}
The functional derivative of the generating function with respect to the
function $\eta$ yields Eq.~(\ref{fq}),
that is, 
\begin{equation}
  \langle 0 |\bar{q}\Gamma (i\Slash{D}-m_q)
q|0\rangle _{\theta}\propto \frac{\delta
  Z[\eta]_{\theta}}{\delta\eta(0)}\biggr|_{\eta =0}~.
\label{prop}
\end{equation}
Now we replace the integration variable $\bar{q}$ with a new integration
variable $\bar{q}^{\prime}$ as
\begin{equation}
 \bar{q}\to \bar{q}^{\prime}=\bar{q}-\eta\bar{q}\Gamma ~.
\end{equation}
Since this step does not change the integral, then we obtain
\begin{equation}
  Z[\eta]= \int{\cal D}\bar{q}{\cal D}q
\left[{\rm Det}\left\{
\frac{\delta\bar{q}^{\prime}}{\delta\bar{q}}\right\}
\right]^{-1}
\exp\left[
i\int d^4x{\cal L}+{\cal O}(\eta^2)\right]~,
\label{rewrite}
\end{equation}
where the inverse of the Jacobian comes from the transformation of the
measure for the fermionic variable. 

Next, we evaluate the Jacobian in the expression above. Since we are
interested in the first order derivative of the generating function, we
expand the Jacobian in $\eta$ and keep only terms linear in $\eta$. 
\begin{equation}
 \frac{\delta\bar{q}^{\prime}_\beta (y)}{\delta \bar{q}_\alpha (x)}
=\delta_{\alpha\beta}\delta^4(x-y)-\eta(y)\Gamma_{\alpha\beta}\delta^4(x-y)+
{\cal O}(\eta^2)~.
\end{equation}
Using the identity
\begin{equation}
 {\rm Det}M=\exp {\rm Tr}\ln M~,
\end{equation}
we readily obtain the Jacobian as
\begin{equation}
{\rm Det}\left\{
\frac{\delta\bar{q}^{\prime}}{\delta\bar{q}}\right\}
=\exp\left[
-{\rm Tr}(\Gamma)\int d^4x\eta(x)\delta^4(x-x)\right]~.
\label{jacobian}
\end{equation}
It is found that if the trace of the matrix $\Gamma$ is non-zero, the
Jacobian yields a singular factor, while if it vanishes we need careful
treatment for evaluating this term. So, in the following discussion, we
divide $\Gamma$ into two types; one is the term proportional to the unit
matrix and the other is the traceless part. 

First, we consider the case $\Gamma \propto \1$.
Using Eqs.~(\ref{prop}), (\ref{rewrite}), and (\ref{jacobian}), we obtain
the following equation:
\begin{equation}
   \langle 0 |\bar{q}\Gamma (i\Slash{D}-m_q)q|0\rangle _{\theta}
=-i{\rm Tr}(\Gamma)\delta^4(0)~.
\end{equation}
Once you carry out the normal ordering for the composite operator
$\bar{q}\Gamma (i\Slash{D}-m_q)q$, the singular
factor in the right-hand side vanishes. Thus we conclude that
after normal ordering,  
\begin{equation}
 \langle 0 |\bar{q}\Gamma(i\Slash{D}-m_q)q|0\rangle _{\theta} =0~,
\label{eom_cond}
\end{equation}
when $\Gamma\propto \1$.  This equation implies that we may use the
equations of motion for quark condensates in this case.

Next, we shall turn to the traceless part. In this case the 
Jacobian in Eq.~(\ref{jacobian}) is written in terms of the anomaly
function defined as
\begin{eqnarray}
{\cal A}(x)&\equiv& 2{\rm Tr}(\Gamma)\delta^4(x-x) .
\label{A_func}
\end{eqnarray}
With this function, Eq.~(\ref{jacobian}) leads to
\begin{equation}
 {\rm Det}^{-1}\left\{
\frac{\delta\bar{q}^{\prime}}{\delta\bar{q}}\right\}
=\exp\left[+\frac{1}{2}
\int d^4x\eta(x){\cal A}(x)\right]~.
\label{det_a}
\end{equation}
The usual analysis for the chiral anomaly tells us that
the function ${\cal A}(x)$ in Eq.~(\ref{A_func}) does not vanish only
for the case $\Gamma=\gamma_5$. Thus, if $\Gamma\neq \gamma_5$,
Eq.~(\ref{eom_cond}) is satisfied.
When $\Gamma=\gamma_5$, on the other hand, the function ${\cal A}(x)$ is
evaluated as 
\begin{equation}
 {\cal A}(x)=\frac{\alpha_s}{4\pi}G^A_{\mu\nu}\tilde{G}^{A\mu\nu}~.
\end{equation}
From Eqs.~(\ref{prop}), (\ref{rewrite}), and (\ref{det_a}) we eventually
find that
\begin{equation}
    \langle 0 |\bar{q}\gamma_5(i\Slash{D}-m_q)q|0\rangle _{\theta}=
-\frac{i\alpha_s}{8\pi}
\langle 0|
G^A_{\mu\nu}
\tilde{G}^{A\mu\nu}|0\rangle_{\theta}~.
\label{resu}
\end{equation}
This expression is simplified via the axial current anomaly equation:
\begin{equation}
 \partial^\mu (\bar{q}\gamma_\mu \gamma_5 q)=
2im_q\bar{q}\gamma_5q
+\frac{\alpha_s}{4\pi}G^A_{\mu\nu}\tilde{G}^{A\mu\nu}~.
\end{equation}
Then, Eq.~(\ref{resu}) leads to 
\begin{equation}
 \langle 0| \bar{q}\gamma_5 i\Slash{D}q|0\rangle_{\theta}
=\frac{1}{2i}\langle 0|\partial^\mu (\bar{q}\gamma_\mu \gamma_5 q)
|0\rangle _{\theta}=0~.
\label{totder}
\end{equation}
Therefore, we are not able to use the classical equations of motion for quark
condensates in this case.

As a result, we find that
\begin{align}
 &\langle 0| \bar{q}\Gamma i\Slash{D}q|0\rangle_{\theta}
\nonumber \\
=&
\begin{cases}
 \langle 0 |\bar{q}\Gamma m_qq|0\rangle _{\theta} & 
({\rm for~}\Gamma=\1, \gamma_\mu,
 \gamma_\mu \gamma_5, \sigma_{\mu\nu}) \\
0 & ({\rm for~}\Gamma=\gamma_5)
\end{cases}~.
\end{align}
Also, its conjugate leads to
\begin{align}
 &-\langle 0| \bar{q}i\overleftarrow{\Slash{D}}\Gamma q|0\rangle_{\theta}
\nonumber \\
=&
\begin{cases}
 \langle 0 |\bar{q}m_q\Gamma q|0\rangle _{\theta} & ({\rm for~}\Gamma=\1, \gamma_\mu,
 \gamma_\mu \gamma_5, \sigma_{\mu\nu}) \\
0 & ({\rm for~}\Gamma=\gamma_5)
\end{cases}~.
\end{align}

Before concluding the section, we add a comment on the condensate of
the total derivative terms. As we have already conducted in
Eq.~(\ref{totder}), the
Lorentz invariance of vacuum implies that condensates of the divergence
of quark bi-linear always vanish, {\it i.e.}, 
\begin{equation}
 \langle 0| \partial^\mu (\bar{q} \Gamma_\mu q)|0\rangle
=   \partial^\mu\langle 0| (\bar{q} \Gamma_\mu q)|0\rangle=0~,
\end{equation} 
with $\Gamma_\mu$ a constant matrix which transforms as a vector under
the Lorentz transformation, such as $\gamma_\mu$, $\gamma_\mu\gamma_5$,
or so. On the other hand, the total derivative of the quark bi-linear is
written as
\begin{equation}
 \partial^\mu (\bar{q} \Gamma_\mu q)= (\partial^\mu\bar{q}) \Gamma_\mu q
+\bar{q} \Gamma_\mu (\partial^\mu q)
=\bar{q}\overleftarrow{D}^\mu \Gamma_\mu q
+\bar{q} \Gamma_\mu D^\mu q~.
\end{equation}
Thus we find
\begin{equation}
 \langle 0|\bar{q}\overleftarrow{D}^\mu \Gamma_\mu q|0\rangle
=-\langle \bar{q} \Gamma_\mu D^\mu q |0\rangle~.
\end{equation}

\section{Wilson Line in Fock-Schwinger gauge}
\label{wilson_line}

Quark fields $q(x)$ are always accompanied by an appropriate Wilson-line
operator in order to compensate the different gauge transformation
property of the quark fields at different space-time points. In the
Fock-Schwinger gauge, however, one may always choose a particular path
which makes the Wilson-line operator equal to
identity \cite{CEBAF-TH-91-11}. We show this statement in the
following. The Wilson line is written as
\begin{equation}
 U_P(x,0)=P\left\{
\exp\left[
ig_s\int_{0}^{1}ds\frac{dx^{\prime \mu}(s)}{ds}G^A_{\mu}(x^{\prime}(s))T^A
\right]
\right\},
\end{equation}
where
\begin{equation}
 x^{\prime}(0)=0,~~~~~~x^{\prime}(1)=x~,
\end{equation}
and $P$ denotes path-ordering. This operator depends on the choice of
the integration path. Here we take a path such that
\begin{equation}
 x^{\prime}(s)=sx~.
\end{equation}
Then,
\begin{equation}
 U_P(x,0)=P\left\{
\exp\left[
ig_s\int_{0}^{1}dsx^\mu G^A_{\mu}(sx)T^A
\right]
\right\}~.
\label{wilson_fs}
\end{equation}
In the Fock-Schwinger field, the gluon field is expanded as
\begin{equation}
 G_\mu(x)=
\frac{1}{2\cdot 0 !} x^{\nu} G_{\nu\mu}(0)
+
\frac{1}{3\cdot 1 !} 
x^\alpha x^{\nu} (D_\alpha  G_{\nu\mu}(0))
+
\frac{1}{4\cdot 2 !} 
x^\alpha x^\beta x^{\nu} (D_\alpha D_\beta G_{\nu\mu}(0))
+
\cdots~.
\end{equation}
Inserting this expression into Eq.~(\ref{wilson_fs}), we readily find
that all terms in the exponential vanish due to the antisymmetric
property of the gluon field strength tensor. Therefore,
\begin{equation}
 U_P(x,0)=1~.
\end{equation}

{}

\end{document}